\documentclass[journal]{IEEEtran}
\usepackage{indentfirst,amsmath}
\usepackage{color}
\usepackage{amsfonts,amsmath,amssymb,amsthm}
\usepackage{latexsym,amscd}
\usepackage{amsbsy}
\usepackage{graphicx,multirow,bm}

\newtheorem{Theorem}{Theorem}

\newtheorem{Example}{Example}

\newtheorem{Remark}{Remark}
\newtheorem{Lemma}{Lemma}

\newtheorem{Construction}{Construction}
\newtheorem{Proposition}{Proposition}
\newcommand{\xiaowuhao}{\fontsize{9pt}{\baselineskip}\selectfont}

\begin{document}

\title{A Framework of Constructions of Minimal Storage Regenerating Codes with the Optimal Access/Update Property
\author{Jie Li, Xiaohu Tang, \IEEEmembership{Member,~IEEE} and Udaya Parampalli, \IEEEmembership{Senior Member,~IEEE}}
\thanks{Manuscript received November 19, 2013; revised November 23, 2014;
accepted February 9, 2015. J. Li and X. Tang were supported in part by the
National Science Foundation of China under Grant 61325005 and in part by
the Major Frontier Project of Sichuan Province. U. Parampalli was supported
in part by the Communications Sensing and Coding Research Network, in
part by the International Research and Research Training Fund, and in part
by the University of Melbourne, Melbourne, VIC, Australia.}
\thanks{J. Li is with the Information Security and National Computing Grid Laboratory, Southwest Jiaotong University, Chengdu, 610031, China (e-mail: jieli873@gmail.com).}
\thanks{X.H. Tang is with the Information Security and National Computing Grid
Laboratory, Southwest Jiaotong University, Chengdu 610031, China, and also
with the Beijing Center for Mathematics and Information Interdisciplinary
Sciences, Beijing 100048, China (e-mail: xhutang@swjtu.edu.cn).}
\thanks{U. Parampalli is with the Department of Computer Science and Software Engineering,
University of Melbourne, VIC 3010, Australia (email:
udaya@unimelb.edu.au).}
}

\date{}
\maketitle

\begin{abstract}
In this paper, we present a generic framework for constructing systematic minimum storage regenerating codes with two parity nodes based on the invariant subspace technique.
Codes constructed in our framework not only contain some best known codes as
special cases, but also include some new codes with key properties such as the optimal access property and the optimal update property. In particular, for a given storage capacity of an individual node, one of the new codes has the largest number of
systematic nodes   and two of the new codes  have the largest number of
systematic nodes with the optimal update property.
\end{abstract}

\begin{IEEEkeywords}
Distributed storage, high rate, invariant subspace, MSR code, optimal access, optimal update.
\end{IEEEkeywords}

\section{Introduction}

\IEEEPARstart{D}{istributed} storage systems with high reliability have wide
applications in large data centers, peer-to-peer storage systems
such as OceanStore \cite{ocean}, Total Recall \cite{total}, DHash++
\cite{Dhash}, and storage in wireless networks. To ensure
reliability,  the redundancy is crucial for these systems. A
popular option to add redundancy is to employ erasure codes which
can efficiently store data and protect against node failures. Examples of
several distributed storage systems that employ erasure codes
are Facebook's coded Hadoop, Google Colossus and Microsoft Azure
\cite{Micro}.

Recently, a new class of erasure codes for distributed storage systems called \emph{minimum storage regenerating} (MSR) codes was introduced in \cite{Dimakis}.
Consider a file of size $\mathcal{M} = k\alpha$ symbols stored across a distributed storage system with $n$
nodes, each keeping $\alpha$ symbols, that deploys an MSR code by
storing the source data on the first $k$ nodes, called \emph{systematic nodes}, and
mixtures of the source data on the other
$n-k$ nodes, termed \emph{parity nodes}.  To provide reliability, MSR codes
must possess two abilities:
\begin{itemize}
\item [(a)] Reconstruction ability: In particular, an MSR code has the \emph{MDS property} that any $k$ out of the $n$ nodes suffice to reconstruct the whole source data.
\item [(b)] Repair  ability: In practical distributed storage systems, the most common failure
is failure of a single node. For this scenario, to maintain redundancy one has to
repair the failed node by downloading $\beta\le \alpha$ symbols from each of any $d\ge k$ surviving nodes.
The \emph{repair bandwidth} $\gamma$ is defined as the amount of data  downloaded during the repair procedure, i.e.,
$\gamma=d\beta$.   In \cite{Dimakis},  MSR codes are shown to have the optimal repair property for the following values:
\begin{equation}\label{MSR}
  \left(\alpha,\gamma\right)=\left(\frac{\mathcal{M}}{k},\frac{\mathcal{M}d}{k(d-k+1)}\right).
\end{equation}
\end{itemize}

Up to now, constructions of MSR codes have attracted a lot of attention \cite{polynomial,Cadambe13,subspace,CDH,hadamard,product,SRKR,SRKR1,Suh,Zigzag,Long,WD09}. However, many constructions have strict constraints on
the parameters $n,k,d$. For example, $d\ge 2k-2$ in \cite{product,SRKR,SRKR1,Suh}, which corresponds to \emph{low rate} (i.e., $\frac{k}{n}\le \frac{1}{2}$) regime. For \emph{high rate} (i.e., $\frac{k}{n}\ge \frac{1}{2}$) regime, most known constructions are built on the concept of interference alignment, which was originally introduced in the context of wireless communication
networks \cite{IA1,IA2}, and was later exploited
for distributed storage systems \cite{WD09}.

In contrast to other known constructions of high rate MSR codes,
the Zigzag code proposed by Tamo \emph{et al.} \cite{Zigzag}
is an MSR code exhibiting two additional interesting properties: the optimal
access property and the optimal update property, which either does not need computing during
the download phase of repair or minimizes the reading/writing during update.
The Zigzag code works for
arbitrary parameters $n,k$ and $d=n-1$, and requires a small finite field size $q$, for example, $q=3$ for $n-k=2$.
It seems that the only
shortcoming of the Zigzag code is the storage $\alpha$  of individual nodes, i.e., $\alpha=r^{k-1}$ grows sharply with the
increase of $k$ where $r=n-k$. In parallel to \cite{Zigzag}, the construction of the Zigzag code has also been discovered by Cadambe \emph{et al.} in \cite{subspace} via subspace interference alignment.
In \cite{Long},
Wang \emph{et al.} presented another MSR code, named long MDS code, that increases
the number $k$ of the systematic nodes to
nearly three times  that of the Zigzag code but still maintains two parity nodes and the same node capacity $\alpha$. However, a larger finite field size is required and none of the systematic
nodes possess the optimal access  property and the optimal update property simultaneously.

In the literature, there are mainly two repair types: exact repair and functional repair.
Compared with the latter, exact repair is preferred since it
does not incur additional significant system overhead by
regenerating the exact replicas of the lost data in the failed node \cite{survey}.
Unfortunately,  except for the one in \cite{hadamard}, all the known  MSR codes
of high rate, including the aforementioned Zigzag code and long MDS code,
can only exactly repair all the systematic nodes optimally
with respect to the bound in \eqref{MSR}, whereas repair the parity node trivially by downloading the whole
original file from all the systematic nodes. For simplicity, throughout this paper we say that such MSR codes possess the
optimal repair property and omit that the property is only valid for systematic nodes.
It should be noted that this kind of code is acceptable for a practical system due to two aspects: (1) The number of parity nodes is quite smaller compared to that of systematic nodes; (2) The failures of systematic nodes and  parity nodes are different since the omission  of some raw information would affect the information access time
for the former \cite{Zigzag}.

In this paper, we focus on  high rate MSR codes. Obviously,  high
rate implies a large value of $k$ for fixed $n$. When $k=n-1$,  the repair bandwidth is the highest, i.e., $\gamma=\mathcal{M}$ by  \eqref{MSR}. Then,
when $k=n-2>1$ and $d=n-1$ (which can reduce the repair bandwidth since $\gamma$ is a decreasing function of $d$ in \eqref{MSR}), MSR codes are of great interest because they can achieve the highest rate $k\over k+2$ for $\gamma=(k+1)\alpha/2<\mathcal{M}$.  Thus, it is very desirable to construct MSR codes with two parity nodes for arbitrary number of systematic
nodes $k$.

The main contribution of this paper
is to present a simple but generic framework to construct MSR codes
with two parity nodes based on the invariant subspace technique. Our
construction not only contains the modified Zigzag code
(the code obtained from the Zigzag code \cite{Zigzag} by deleting its first node), and the long MDS code \cite{Long} as
special cases, but also generates some new MSR codes. Specifically,
based on the modified Zigzag code with $m$ systematic nodes, we can obtain three new MSR codes by adding
$2m$ or $m$ more systematic nodes. When adding $2m$ more systematic nodes without the optimal
access property and the optimal update property, we can construct new code  $\mathcal{C}_1$ over a finite field
of size $q\ge 2m+1$. When adding $m$ more systematic nodes,we can make a choice of either  a smaller finite field or new
nodes having the optimal update complexity. For the former, the finite
field size can be reduced to $q\ge m+1$, which
results in new code $\mathcal{C}_2$. For the latter, the
resulting new code  $\mathcal{C}_3$ still requires a finite field
size $q\ge 2m+1$. In addition, another new code $\mathcal{C}_4$
which has the same number of systematic nodes and requires the same size of finite field as those of $\mathcal{C}_2$ can be derived. All the systematic nodes of $\mathcal{C}_4$ have the optimal update property but none of them have the optimal access property. In this sense, we provide four code constructions with different parameters that allows for trading-off between
 the size of the finite field and the number of systematic
nodes (with the optimal access/update property). In particular, given an $\alpha$, the code
$\mathcal{C}_1$ has the largest number of
systematic nodes, while $\mathcal{C}_3$ and $\mathcal{C}_4$ have the largest number of
systematic nodes with the optimal update property.  For
comparison, the parameters of the new codes, the Zigzag code, and the long
MDS code are listed in Table \ref{table para compare}.
\begin{table*}[htbp]\label{table para compare}
\centering
\caption{Comparison between the new codes and some known codes with two parity nodes and $\alpha=2^m$,
 where $k$, $k_A$, $k_U$ and $k_{A\&U}$ denote the number of systematic nodes, the number of systematic nodes with the optimal access property,  the number of systematic nodes with the optimal update property and the number of systematic nodes with both the optimal access property and the optimal update property respectively, and $q$ denotes the size of the finite field required.
}
\begin{tabular}{|c|c|c|c|c|c|c|}
\hline &New&New&New&New&The Zigzag&The Long MDS\\
&code $\mathcal{C}_1$& code $\mathcal{C}_2$& code $\mathcal{C}_3$& code $\mathcal{C}_4$& code \cite{Zigzag}&code \cite{Long}\\
\hline $k$&$3m$&$2m$&$2m$&$2m$&$m+1$&$3m$\\
\hline $k_A$&$m$&$m$&$m$&$0$&$m+1$&$2m$\\
\hline $k_U$&$m$&$m$&$2m$ &$2m$&$m+1$&$m$\\
\hline $k_{A\&U}$&$m$&$m$&$m$ &$0$&$m+1$&$0$\\
\hline $q$&$\geq 2m+1$&$\geq m+1$&$\geq 2m+1$&$\geq m+1$&$3$&$\geq 2m+1$\\
\hline
\end{tabular}
\end{table*}

The rest of this paper is organized as follows. Section II gives preliminaries about the necessary and sufficient conditions for an erasure code with two parity nodes to be an MSR code, and presents the special partition for a given basis.  Section III proposes the generic construction, by which some known codes are reinterpreted and four new MSR codes with the optimal access/update property are  derived.  Finally,  Section IV
draws concluding remarks.

\section{Preliminaries}

Let $q$ be a prime power, $\mathbf{F}_q$ be the finite field with $q$ elements, and $\mathbf{F}_q^{l}$ be the vector space of dimension $l$ over $\mathbf{F}_q$.
For simplicity, throughout this paper we do not specifically distinguish the vector space
spanned by row vectors or column vectors if the context is clear.

Assume  that a file of size $\mathcal{M}=k\alpha$ denoted by
the column vector  $f\in\mathbf{F}_q^{k\alpha}$ is partitioned in $k$ parts $f=[f_1^T f_2^T \cdots f_k^T]^T$, each of size $\alpha$,
where $T$ denotes the transpose operator. We encode $f$ using an $(n=k+2,k)$  MSR code $\mathcal{C}$ and  store it
across $k$ systematic and
two parity storage nodes. Precisely, the first $k$  (systematic) nodes store the file parts $f_1,f_2,\cdots,f_k$ in an uncoded form respectively,
and the parity nodes store linear combinations of $f_1,f_2,\cdots, f_k$. Without loss of generality, it is assumed that
the nodes $k+1$ and $k+2$ respectively store $f_{k+1}=f_1+f_2+\cdots+f_k$  and $f_{k+2}=\sum\limits_{i=1}^{k}A_{i}f_i$ for some $\alpha\times\alpha$ matrices $A_{1},\cdots,A_k$ over $\mathbf{F}_q$, where the matrix $A_{i}$ is called the \emph{coding matrix} for the $i$th systematic node, $1\le i\le k$. Table \ref{MSR_Model} illustrates the structure of a $(k+2,k)$ MSR code.

\begin{table}[htbp]\label{MSR_Model}
\begin{center}
\caption{Structure of a $(k+2,k)$ MSR code}
\begin{tabular}{|c|c|}
\hline
Systematic node & Systematic data \\
\hline
1 & $f_1$ \\
\hline
\vdots & \vdots \\
\hline
$k$ & $f_k$ \\
\hline
Parity node & Parity data \\
\hline
1 & $f_{k+1}=f_1+ \cdots+ f_k$ \\
\hline
2 & $f_{k+2}=A_1f_1+\cdots + A_kf_k$ \\
\hline
\end{tabular}
\end{center}
\end{table}

Note that reconstruction of the original file demands that (i) $A_i$ is invertible when
connecting nodes belong to the set $\{1,2,\cdots,k+1\}\backslash\{i\}$ (or $\{1,2,\cdots,k,k+2\}\backslash\{i\}$), for any  $1\le i\le k$ and (ii) $A_i-A_j$ is invertible
when connecting nodes belong to the set $\{1,2,\cdots,k+2\}\backslash\{i,j\}$, for any $1\le i\ne j\le k$.
In other words, the MSR code $\mathcal{C}$ with the MDS property requires  \cite{Long}
\begin{itemize}
\item [R1.]  $A_i$ and $A_i-A_j$ are all invertible for any $1\le j\neq i\le k$.
\end{itemize}

As mentioned in the last section, $d$ is assumed to be $n-1$ for minimizing the repair bandwidth.
Then in order to repair a failed node, only half of data is downloaded from each surviving node.  When a systematic node $i$ fails, we download data $S_{i,j}f_j$ from node $j\neq i$ using an $\frac{\alpha}{2}\times\alpha$ matrix  $S_{i,j}$ of rank $\frac{\alpha}{2}$, where $S_{i,j}$ is referred to as the \emph{repair matrix} of the $j$th node for the $i$th systematic node.
To simplify the repair strategy, we assume $S_{i,j}=S_i$ for all $1\le i\le k$, $1\le j\ne i\le k+2$. Then during the repair process of a node $i$,
one downloads $S_if_j$ from each node $1\le j\ne i\le k+2$, and eventually gets
the following system of linear equations
\begin{eqnarray*}\label{repair}
\left(\begin{array}{c}
S_i f_{k+1}\\
S_i f_{k+2}
\end{array}\right)=\underbrace{\left(\begin{array}{c}
S_i\\
S_i A_i
\end{array}\right)f_i}_{\mathrm{useful~ data}}+\sum_{j=1,j\ne i}^{k}\underbrace{\left(\begin{array}{c}
S_i\\
S_i A_j
\end{array}\right)f_j}_{\mathrm{interference~ by}~f_{j}}.
\end{eqnarray*}
\begin{Remark}
 A  $(k+2,k)$ MSR code with $f_{k+1}=f_1+f_2+\cdots+f_k$ and $S_{i,j}=S_i$  can be viewed as a kind of canonical form \cite{subspace,hadamard,Zigzag,Access,Long}. Firstly, if $f_{k+1}=B_1 f_1+B_2 f_2+\cdots+B_k f_k$ for some nonsingular $\alpha*\alpha$ matrices
 $B_j$, $1\le j\le k$,  then the code can be equivalently converted to the following code
\begin{center}
\begin{tabular}{|c|c|}
\hline
Systematic node & Systematic data \\
\hline
1 & $f_1'$ \\
\hline
\vdots & \vdots \\
\hline
$k$ & $f_k'$ \\
\hline
Parity node & Parity data \\
\hline
1 & $f_{k+1}'=f_1'+ \cdots+ f_k'$ \\
\hline
2 & $f_{k+2}'=A_1'f_1'+\cdots + A_k'f_k'$ \\
\hline
\end{tabular}
\end{center}
where $f_i'=B_i f_i$ and $A_i'=A_iB_i^{-1}$ for any $1\le i\le k$ by using
repair matrices $S_{i,j}'=S_{i,j}B_j^{-1}, 1\le j\ne i\le k, S_{i,k+1}'=S_{i,k+1}$ and $S_{i,k+2}'=S_{i,k+2}$. Secondly,
 as shown in \cite{Access}, such a  $(k+2,k)$ MSR code can be transformed to a $(k+1,k-1)$ MSR code in canonical form. Thus we only consider MSR codes in canonical form since  the difference between the numbers of their nodes $k+2$ and $k+1$  is  negligible.
\end{Remark}
Then, the optimal repair property needs  to cancel  all the interference terms by R2 and then recover
the original data $f_i$ by R3 \cite{Long}:
\begin{itemize}
\item [R2.] $\mbox{rank}\left(\left(
                 \begin{array}{c}
                   S_{i} \\
                   S_{i}A_{j} \\
                 \end{array}
               \right)\right)=\frac{\alpha}{2}
$  for any $1\le j\neq i\le k$.
\vspace{2mm}
\item [R3.] $\mbox{rank}\left(\left(
                 \begin{array}{c}
                   S_{i}\\
                   S_{i}A_{i}
                 \end{array}
               \right)\right)=\alpha$ for all $1\le  i\le k$.

\end{itemize}

The repair procedure firstly  computes $S_if_j$, $1\le j\ne i\le k+2$, and then transmits the result
to the newcomer storage node. A systematic node is said to have the \emph{optimal access property} if the computation within the surviving nodes is not required during the repair procedure \cite{Long}.
For some applications such as data centers, the access to information is more costly than the transmission,
which may cause a bottleneck if  the amount of the former is larger than that of latter \cite{Access}. Hence, an MSR code with more systematic nodes possessing the optimal access property is preferred. It is easy to verify that
the $i$th systematic node with the optimal access property requires
\begin{itemize}
\item [R4.] Each row of $S_i$ has only one nonzero element, which equals to $1$.
\end{itemize}
In addition, when a symbol in a systematic node is rewritten, if only the symbol itself and one symbol at each parity node need an update, then the systematic node is said to have the \emph{optimal update property}, which achieves the minimum reading/writing during writing of information \cite{Zigzag}.  Therefore, an MSR code with more systematic nodes possessing the optimal update property is desired especially in a system where updates are frequent.
In fact, the $i$th systematic node with the optimal update property is equivalent to that every parity element is a linear combination of exactly one element from the $i$th systematic node, i.e.,
\begin{itemize}
\item [R5.] Each column of $A_i$ has only one nonzero element.
\end{itemize}

Usually, it is favorable for a code to have more systematic nodes for a given $\alpha$. Recall that the number $k$ of systematic nodes of
the Zigzag code is much less than that of the long MDS code.
In this paper, we therefore mainly aim
at increasing $k$ of the Zigzag code. According to R1, R4 and R5, however, a systematic node has the optimal update property if and only if its coding matrix $A_i$ is either a diagonal matrix or product of a diagonal matrix and a permutation matrix; a systematic node has the optimal access property if and only if its repair matrix $S_i$ is an ${\alpha\over 2}\times \alpha$ submatrix of an $\alpha\times \alpha$ permutation matrix. The number of distinct such matrices satisfying R2 and R3 appears to be greatly limited. In \cite{Zigzag,Access}, it is shown that the largest number of systematic nodes of an MSR code with the optimal access property (resp. both the optimal access property and the optimal update property) is $2\log_2\alpha$ (resp. $\log_2\alpha+1$).

In what follows, we introduce two useful tools: invariant subspaces and partition sets, which enable us to  construct our generic coding matrices and repair matrices satisfying R2 and R3.

\subsection{Invariant subspaces}

In this subsection, we determine the coding matrices by using invariant subspaces.

For a matrix $A$,  denote by span$(A)$ the vector space spanned by its rows, obviously $\mbox{dim}(\mbox{span}(A))=\mbox{rank}(A)$.
Recall that $S_i$ is a matrix of rank ${\alpha\over 2}$. Then,  R2 implies that $\mbox{span}(S_i A_{j})\subseteq\mbox{span}(S_i)$.
Moreover, it follows from R1 that $A_{j}$ is of full rank $\alpha$
and consequently we have $\mbox{rank}(S_i A_{j})=\mbox{rank}(S_i)$. Hence,
$\mbox{dim}(\mbox{span}(S_i A_{j}))=\mbox{dim}(\mbox{span}(S_i))$, i.e.,
\begin{eqnarray}\label{Eq_IVS}
\mbox{span}(S_i A_{j})=\mbox{span}(S_i)
\end{eqnarray}
which indicates that $\mbox{span}(S_i)$ is an \emph{invariant subspace} of vector space $\mbox{span}(A_{j})=\mathbf{F}_q^{\alpha}$ with respect to
the linear transformation $T$ defined by
\begin{equation}\label{Eq_LT}
T(x)=xA_{j},\ \ \mbox{for\ any\ }x\in \mathbf{F}_q^{\alpha}.
\end{equation}

Firstly let us look at a simple example.  Let $S=\left(
                                    \begin{array}{c}
                                      e_0 \\
                                      e_1
                                    \end{array}
                                  \right)$
where $e_0,e_1$ are two arbitrary row vectors of length $\alpha$ over $\mathbf{F}_q$, and they are linearly independent.
Then by \eqref{Eq_IVS},  span$(S)$ is an invariant subspace of span$(A)$ with respect to $T: x\mapsto xA$ if and only if
                                  \begin{equation*}
                                    \left(
                                    \begin{array}{c}
                                      e_0 \\
                                      e_1 \\
                                    \end{array}
                                  \right)A=\left(\begin{array}{c}
                                      ae_0+be_1 \\
                                      ce_0+de_1 \\
                                    \end{array}
                                  \right)\ \mbox{and}\ ad\neq bc,\ a,b,c,d\in \mathbf{F}_q.
                                  \end{equation*}
In details, there are 7 cases as below:
\begin{itemize}
\item [] Case 1: $b=c=0$ and $a,d\neq0$,
\vspace{1mm}
\item [] Case 2: $a=d=0$ and $b,c\neq0$,
\vspace{1mm}
\item [] Case 3: $b=0$ and $a,c,d\neq0$,
\vspace{1mm}
\item [] Case 4: $a=0$ and $b,c,d\neq0$,
\vspace{1mm}
\item []Case 5: $a,b,c,d\neq0$ and $ad\neq bc$,
\vspace{1mm}
\item [] Case 6: $c=0$ and $a,b,d\neq0$,
\vspace{1mm}
\item [] Case 7: $d=0$ and $a,b,c\neq0$.
\end{itemize}
Note that if  we interchange $e_0$ with $e_1$, Case 3 (respectively, 4) will become Case 6 (respectively, 7).
Besides, the coding matrix corresponding to Case 5 is a summation of two coding matrices corresponding to Cases 3 and 4,  which
would cause higher update complexity for its corresponding systematic node than that for the latter two.
Therefore, we mainly consider Cases 1-4.
Specifically, we say that the pair  $(e_0,e_1)$ with respect to $A$ is
\begin{itemize}
  \item type I  if $\left(
\begin{array}{c}
e_0 \\
e_1 \\
\end{array}
\right)A=\left(\begin{array}{c}
ae_0 \\
de_1 \\
\end{array}
\right)$,
\vspace{2mm}
  \item  type II  if $\left(
\begin{array}{c}
e_0 \\
e_1 \\
\end{array}
\right)A=\left(\begin{array}{c}
be_1 \\
ce_0 \\
\end{array}
\right)$,
\vspace{2mm}
  \item  type III  if $\left(
\begin{array}{c}
e_0 \\
e_1 \\
\end{array}
\right)A=\left(\begin{array}{c}
ae_0 \\
ce_0+de_1 \\
\end{array}
\right)$,
\vspace{2mm}
  \item  type IV if $\left(
\begin{array}{c}
e_0 \\
e_1 \\
\end{array}
\right)A=\left(\begin{array}{c}
be_1 \\
ce_0+de_1 \\
\end{array}
\right)$.
\end{itemize}

Now we extend the analysis to the general case. From now on, let $\{e_0,\cdots,e_{2^m-1}\}$ be the standard basis of $\mathbf{F}_q^{\alpha}$ where $\alpha=2^m$, i.e.,
$i$th basis vector
\begin{equation*}
    e_i=(0,\cdots,0,1,0,\cdots,0),\,\,0\le i\le 2^m-1,
\end{equation*}
with only the $i$th entry being nonzero. Divide the basis into $2^{m-1}$ pairs, i.e.,
\begin{equation}\label{B4}
    (e_{i_1},e_{j_1}),(e_{i_2},e_{j_2}),\cdots,(e_{i_{2^{m-1}}},e_{j_{2^{m-1}}}),
\end{equation}
where $0\le i_1<i_2<\cdots<i_{2^{m-1}}\le 2^m-1$, $0\le j_1<j_2<\cdots<j_{2^{m-1}}\le 2^m-1$ and $i_s\ne j_t$ for any $1\le s,t\le 2^{m-1}$.
For simplicity, assume that
any pair forms an invariant subspace of $\mathbf{F}_q^{\alpha}$ with respect to $T$ and all the
pairs are of the same type, i.e.,
\begin{eqnarray*}
                                    \left(
                                    \begin{array}{c}
                                      e_{i_1} \\
                                      \vdots\\
                                      e_{i_{2^{m-1}}} \\
                                      e_{j_1} \\
                                      \vdots\\
                                      e_{j_{2^{m-1}}}
                                    \end{array}
                                  \right)A=\left(\begin{array}{c}
                                      a_{i_1} e_{i_1}+ b_{j_1} e_{j_1} \\
                                      \vdots\\
                                      a_{i_{2^{m-1}}} e_{i_{2^{m-1}}}+b_{j_{2^{m-1}}} e_{j_{2^{m-1}}} \\
                                      c_{i_1} e_{i_1} + d_{j_1} e_{j_1}\\
                                      \vdots\\
                                      c_{i_{2^{m-1}}} e_{i_{2^{m-1}}}+d_{j_{2^{m-1}}} e_{j_{2^{m-1}}}
                                  \end{array}
                                  \right)
\end{eqnarray*}
where $a_i,b_i,c_i$ and $d_i$ are some constants, then the coding matrix $A$ can be uniquely determined. Accordingly, we call $A$ type I, II, III, IV coding matrix respectively.
By convenience, write
\begin{eqnarray*}
                                    \left(
                                    \begin{array}{c}
                                    V_0\\
                                    V_1
                                    \end{array}
                                  \right)A=\left(\begin{array}{c}
                                      a V_0+ b V_1 \\
                                      c V_0 + d V_1
                                  \end{array}
                                  \right)
\end{eqnarray*}
where $a$, $b$, $c$ and $d$ can be coefficients in $\mathbf{F}_q$ or $\frac{\alpha}{2}\times \frac{\alpha}{2}$ diagonal matrices over $\mathbf{F}_q$ and
\begin{eqnarray}\label{Eqn_SetV}
V_0=\left(
                                    \begin{array}{c}
                                      e_{i_1} \\
                                      \vdots\\
                                      e_{i_{2^{m-1}}}
                                    \end{array}
                                  \right), ~V_1=\left(
                                    \begin{array}{c}
                                      e_{j_1} \\
                                      \vdots\\
                                      e_{j_{2^{m-1}}}
                                    \end{array}
                                  \right),
\end{eqnarray}
and still use $V_0$ and $V_1$ to represent their corresponding sets $\{e_{i_1},e_{i_2},\cdots,e_{i_{2^{m-1}}}\}$ and $\{e_{j_1},e_{j_2},\cdots,e_{j_{2^{m-1}}}\}$ respectively in the following
sections if the context is clear.

\subsection{Partition of the basis $\{e_0,\cdots,e_{2^m-1}\}$}

In this subsection, we present a class of partition sets of the basis of $\mathbf{F}_q^{\alpha}$ to obtain  $V_0$ and $V_1$
in \eqref{Eqn_SetV} , which had been used in \cite{Long}, and will be crucial to our constructions as well.

Assume that there are $m$  partition sets of the basis of $\mathbf{F}_q^{\alpha}$ as follows
\begin{equation}\label{A3}
    \{e_0,e_1,\cdots,e_{2^m-1}\}=V_{1,0}\cup V_{1,1}=\cdots=V_{m,0}\cup V_{m,1}
\end{equation}  such that
\begin{equation}\label{A2}
    |V_{i_1,j_1}\cap V_{i_2,j_2}\cap\cdots\cap V_{i_l,j_l}|=2^{m-l}
\end{equation}
for any $1\le i_1<i_2<\cdots<i_l\le m$, $j_t=0,1$,  $1\le t\le l\le m$. It should be noted that \eqref{A2} is useful when designing the code satisfying R2 and R3. Clearly, $|V_{1,j_1}\cap V_{2,j_2}\cap\cdots\cap V_{m,j_m}|=1$ for any $j_1,j_2,\cdots,j_m\in\{0,1\}$ by \eqref{A2}. Without loss of generality, we can set
\begin{eqnarray*}
\{e_j\}=\{e_{(j_1,j_2,\cdots,j_m)}\}=V_{1,j_1}\cap V_{2,j_2}\cap\cdots\cap V_{m,j_m},
\end{eqnarray*}
where $(j_1,j_2,\cdots,j_m)$ is the binary expansion of the integer $j$.
Recursively applying \eqref{A2} to $l=m-1,\cdots, 1$, we then get
\begin{eqnarray}\label{Eqn_Vt}
V_{i,t}=\{e_j|j_i=t\}
\end{eqnarray}
 for $1\le i\le m$ and $t=0,1$. Table III gives two examples of the set partitions that satisfy (\ref{A3}) and (\ref{A2}).

\begin{table}[htbp]\label{example partition}
\begin{center}
\caption{(a) and (b) denote the $m$ set partitions of $V$ that satisfy (\ref{A3}) and (\ref{A2}) for $m=2$ and $m=3$, respectively.}
\begin{tabular}{|c|c|c|c|c|c|}
\hline $i$ & 1 & 2 &$i$ & 1 & 2 \\
\hline \multirow{2}{*}{$V_{i,0}$ }&$e_0$&$e_0$&\multirow{2}{*}{$V_{i,1}$ }&$e_2$&$e_1$\\  & $e_1$&$e_2$&& $e_3$&$e_3$\\
\hline\multicolumn{6}{c}{(A)}
\end{tabular}\vspace{5mm}
\begin{tabular}{|c|c|c|c|c|c|c|c|}
\hline $i$ & 1 & 2 & 3 &$i$ & 1 & 2 & 3\\
\hline \multirow{4}{*}{$V_{i,0}$ }&$e_0$&$e_0$&$e_0$&\multirow{4}{*}{$V_{i,1}$ }&$e_4$&$e_2$&$e_1$\\  & $e_1$&$e_1$&$e_2$&& $e_5$&$e_3$&$e_3$\\&$e_2$&$e_4$&$e_4$&& $e_6$&$e_6$&$e_5$\\&$e_3$&$e_5$&$e_6$&& $e_7$&$e_7$&$e_7$\\
\hline\multicolumn{8}{c}{(B)}
\end{tabular}
\end{center}
\end{table}

Based on the $m$  partition sets in \eqref{Eqn_Vt}, define
\begin{equation}\label{B3}
    V_{i+sm,t}=V_{i,t},\ \  i=1,2,\cdots,m, \ \  s\in \mathbf{N}^*,\ \  t=0,1.
\end{equation}
For any $1\le i_1, i_2\le sm$ and $i_1\not \equiv i_2\mbox{\ mod\ } m$, define $V_{i_1,i_2,j_1,j_2}=V_{i_2,i_1,j_2,j_1}=V_{i_1,j_1}\cap V_{i_2,j_2}$ for $j_1,j_2=0,1$.
Then \begin{eqnarray}\label{Eqn_B3}
\nonumber V_{i_1,j_1}&=&(V_{i_1,j_1}\cap V_{i_2,0})\bigcup (V_{i_1,j_1}\cap V_{i_2,1})\\
&=&V_{i_1,i_2,j_1,0}\cup V_{i_1,i_2,j_1,1},
\end{eqnarray}
and thus we have the following results, which will be frequently used in the sequel.

\begin{Lemma}\label{rank}
For any $i,j\geq1$ and $i\not \equiv j\mbox{\ mod\ } m$, we have
\begin{itemize}
 \item [(i)] \begin{eqnarray*}
 &&\mbox{rank}\left(A_i-A_j\right)\\ &=& \mbox{rank}\left(\left(
                                                              \begin{array}{c}
                                                                V_{i,0} \\
                                                                V_{i,1} \\
                                                              \end{array}
                                                            \right)(A_i-A_j)\right) \\
   &=& \mbox{rank}\left(\left(
           \begin{array}{c}
             V_{i,j,0,0} \\
             V_{i,j,0,1} \\
              V_{i,j,1,0} \\
             V_{i,j,1,1} \\
           \end{array}
         \right)(A_i-A_j)\right),
\end{eqnarray*}

  \item  [(ii)]
\begin{eqnarray*}
   && \mbox{rank}\left(\left(
                         \begin{array}{l}
                           V_{i,0}+u_iV_{i,1} \\
                           (V_{i,0}+u_iV_{i,1})A_j \\
                         \end{array}
                       \right)\right) \\
   &=& \mbox{rank}\left(\left(
                         \begin{array}{l}
                           V_{i,j,0,0}+u_iV_{i,j,1,0} \\
                           V_{i,j,0,1}+u_iV_{i,j,1,1} \\
                           (V_{i,j,0,0}+u_iV_{i,j,1,0})A_j \\
                           (V_{i,j,0,1}+u_iV_{i,j,1,1})A_j \\
                         \end{array}\right)\right)
\end{eqnarray*}
 where $u_i\in \mathbf{F}_q$.
\end{itemize}
\end{Lemma}
\begin{IEEEproof}
The proof is given in Appendix.
\end{IEEEproof}

\section{Generic construction of codes with 2 parity nodes}
In this section, we construct MSR codes with parameters $n=k+2$ and $k=tm$, where $t,m$ are some integers and $\alpha=2^m$, with
the coding matrices being the types defined in subsection 2.1.

\textbf{Generic Construction:}
The $(n=k+2,k)$ code $\mathcal{C}$ has $\alpha\times \alpha$ coding matrices $A_i$  and $\frac{\alpha}{2}\times \alpha$ repair matrices $S_i$  for $1\le i\le k$, such that
\begin{enumerate}
  \item $
    \left(
  \begin{array}{c}
    V_{i,0} \\
    V_{i,1} \\
  \end{array}
\right)A_i=\left(
             \begin{array}{c}
               a_iV_{i,0}+b_iV_{i,1} \\
               c_iV_{i,0}+d_iV_{i,1} \\
             \end{array}
           \right)$ for $1\le i\le k$,
\vspace{2mm}
  \item
  $S_i=V_{i,1}$ or $V_{i,0}+t_{i}V_{i,1}$ for $1\le i\le k$,
\end{enumerate}where $a_i$, $b_i$, $c_i$, $d_i$ and $t_i$ can be coefficients in $\mathbf{F}_q$ or $\frac{\alpha}{2}\times \frac{\alpha}{2}$ diagonal matrices  over $\mathbf{F}_q$ such that $$\left(
             \begin{array}{c}
               a_iV_{i,0}+b_iV_{i,1} \\
               c_iV_{i,0}+d_iV_{i,1} \\
             \end{array}
           \right)$$ is invertible for $1\le i\le k$.

As for Generic Construction, we have the following proposition.

\begin{Proposition}\label{Prop prin}
For a $(k+2,k)$ MSR code  generated by the generic construction,
\begin{itemize}
  \item [(i)] $S_i\ne S_j$ for any $1\le i\ne j\le k$;
  \item [(ii)]There do not exist four repair matrices $S_{j_1}, S_{j_2}, S_{j_3}$ and $S_{j_4}$ such that
  $S_{j_l}=V_{i,0}+t_l V_{i,1}$, $1\le l\le 3$, and $S_{j_4}=V_{i,1}$  or  $V_{i,0}+t_4V_{i,1}$, for an integer $1\le i\le m$ where $j_1,j_2,j_3,j_4$ are four distinct integers
  in $\{1,\cdots,k\}$ and $t_1,t_2,t_3,t_4$ are four distinct elements or matrices over $\mathbf{F}_q$;
\end{itemize}
\end{Proposition}
\begin{IEEEproof}
The proof is given in Appendix.
\end{IEEEproof}

According to Proposition \ref{Prop prin}, in a $(k+2,k)$ MSR code generated by the generic construction, there are at most three repair matrices of the form $S_{l}=V_{i,1}$ or $V_{i,0}+t_l V_{i,1}$,
each appearing at most once, for any given $1\le i\le m$, i.e., the number of systematic nodes is bounded by $k\le 3m$.
In the following, through choosing some appropriate coding matrices in our framework, several $(k+2,k)$ MSR codes, $k\le 3m$,  with  the optimal access property and/or the optimal update property are obtained. This generates not only the known constructions such as the Zigzag code (except for one node) \cite{Zigzag} and the long MDS code \cite{Long},  but also some new codes.

\subsection{Reinterpretation of known constructions}

Based on coding matrices of type II, construct an $(n=k+2,k=m)$  code by
\begin{itemize}
  \item $
    \left(
  \begin{array}{c}
    V_{i,0} \\
    V_{i,1} \\
  \end{array}
\right)A_i=\left(
             \begin{array}{c}
               \Lambda_{i,1}V_{i,1} \\
               \Lambda_{i,0}V_{i,0} \\
             \end{array}
           \right)$ for $1\le i\le m$,
\vspace{2mm}
\item
  $S_i=V_{i,0}$, for $1\le i\le m$,
\end{itemize}
where $\Lambda_{i,0}$ and $\Lambda_{i,1}$ are $\frac{\alpha}{2}\times \frac{\alpha}{2}$ diagonal matrices over $\mathbf{F}_q$.
In fact, it is a modification of the Zigzag code by deleting its first node \cite{Zigzag}. The modified Zigzag code has almost the same properties as that of the Zigzag code, i.e., all the systematic nodes of the modified Zigzag code possess both the optimal access property and the optimal update property.

Through a combination of coding matrices of  types I , III and VI,
the long MDS code \cite{Long} can also be constructed by
\begin{itemize}
  \item
$
  \ \ \   \left(
  \begin{array}{c}
    V_{i,0} \\
    V_{i,1} \\
  \end{array}
\right)A_i\\[4pt]=\left\{\begin{array}{ll}
\left(
               \begin{array}{l}
                 \lambda_{i,0}V_{i,0}+k_iV_{i,1} \\
                  \lambda_{i,1}V_{i,1}\\
               \end{array}
             \right), &\mbox{if}\ \  1\le i\le m\\[12pt]
    \left(
                 \begin{array}{l}
                  \lambda_{i,0} V_{i,0} \\
                \lambda_{i,1}V_{i,1}+k_{i}V_{i,0}\\
                 \end{array}
               \right),&\mbox{if}\ \   m+1\le i\le 2m\\[12pt]
    \left(
                 \begin{array}{l}
                  \lambda_{i,0} V_{i,0} \\
                \lambda_{i,1}V_{i,1}\\
                 \end{array}
               \right),&\mbox{if}\ \   2m+1\le i\le 3m
\end{array}
\right.$
\vspace{2mm}
\item
  $S_i=\left\{\begin{array}{lll}V_{i,0},&\mbox{if}\ \   1\le i\le m
\\V_{i,1},&\mbox{if}\ \   m+1\le i\le 2m
\\
V_{i,0}+V_{i,1},&\mbox{if}\ \  2m+1\le i\le 3m \end{array}\right.$
\end{itemize}
where $\lambda_{i,0},\lambda_{i,1}\in \mathbf{F}_q^*$,  $k_j=\lambda_{j,0}-\lambda_{j,1}$ and $k_{j+m}=\lambda_{j+m,1}-\lambda_{j+m,0}$ for all $1\le i\le k$ and $1\le j\le m$.

Moreover, it is possible to choose $\Lambda_{i,s}$ and $\lambda_{j,s}$ respectively in the constructions of the modified Zigzag code and the long MDS code \cite{Long} such that the conditions R1-R5 are satisfied.

\subsection{New code $\mathcal{C}_1$}

Using the coding matrices of  types II and III, we  construct the first new code.

\begin{Construction}\label{Con_c4}
The $(n=k+2,k=3m)$ code $\mathcal{C}_1$ has $\alpha\times \alpha$ coding matrices $A_i$ and $\frac{\alpha}{2}\times \alpha$ repair matrices $S_i$  for $1\le i\le k$, such that
\begin{enumerate}
\item  $
  \ \ \  \left(
  \begin{array}{c}
    V_{i,0} \\
    V_{i,1} \\
  \end{array}
\right)A_i\\[4pt]=\left\{\begin{array}{ll}
\left(
             \begin{array}{c}
               \lambda_{i,1}V_{i,1} \\
               \lambda_{i,0}V_{i,0} \\
             \end{array}
           \right), &\mbox{if}\ \  1\le i\le m\\[12pt]
    \left(
                 \begin{array}{l}
                  \lambda_{i,0} V_{i,0} \\
                \lambda_{i,1}V_{i,1}+k_{i-m}V_{i,0}\\
                 \end{array}
               \right),&\mbox{if}\ \  m+1\le i\le 3m
\end{array}
\right.$
\vspace{2mm}
\item
  $S_i=\left\{\begin{array}{ll}V_{i,0},&\mbox{if}\ \   1\le i\le m
\\
V_{i,0}+t_{i-m}V_{i,1},&\mbox{if}\ \  m+1\le i\le 3m \end{array}\right.$
\end{enumerate}where $\lambda_{i,0},\lambda_{i,1},k_j,t_j\in \mathbf{F}_q^*$ for all $1\le i\le k$ and $1\le j\le 2m$.
\end{Construction}

\begin{Theorem}\label{Thm_4_1}
$\mathcal{C}_1$ is  a code with the MDS property if and only if
\begin{enumerate}
  \item [(i)] $\lambda_{i,0}\lambda_{i,1}\neq \lambda_{j,0}\lambda_{j,1}$ for any $1\le i\neq j\le m$,
\vspace{2mm}
  \item [(ii)] $\left\{\begin{array}{ll}\lambda_{i,s}\ne \lambda_{j,s}, &\mbox{if}\ \ j=i+m \\ \lambda_{i,s}\ne \lambda_{j,t}, &\mbox{if}\ \ j\neq i+m\end{array}\right.$\\ for any $m+1\le i<j\le 3m$ and $s,t=0,1$,
\vspace{2mm}
  \item [(iii)] $\left\{\begin{array}{ll}\lambda_{i,1}(\lambda_{i,0}-k_{j-m})\neq \lambda_{j,0}\lambda_{j,1}, &\mbox{if}\ \ j=i+m, i+2m \\ \lambda_{i,0}\lambda_{i,1}\neq\lambda_{j,0}^2,\lambda_{j,1}^2, &\mbox{otherwise}\end{array}\right.$\\
for any $1\le i\le m$ and  $m+1\le j\le 3m$.
\end{enumerate}
\end{Theorem}

\begin{IEEEproof}
The proof is given in Appendix.
\end{IEEEproof}

\begin{Theorem}\label{Thm_4_2}
$\mathcal{C}_1$ is a code with the optimal repair property if and only if
\begin{enumerate}
  \item [(i)] $\lambda_{i,1}=t_i^2\lambda_{i,0}$ and $t_i=-t_{i+m}$ for all $1\le i\le m$,
  \item [(ii)] $\lambda_{i,1}=\lambda_{i,0}+t_{i}k_{i-m}$ and $\lambda_{i+m,1}=\lambda_{i+m,0}+t_{i-m}k_{i}$ for all $m+1\le i\le 2m$,
  \item [(iii)] $\mathbf{F}_q$ is of odd characteristic.
\end{enumerate}
\end{Theorem}

\begin{IEEEproof}
The proof is given in Appendix.
\end{IEEEproof}

\begin{Theorem}\label{Thm_4 3}
The first $m$ systematic nodes of $\mathcal{C}_1$  have both the optimal access property and the optimal update property.
\end{Theorem}
\begin{IEEEproof}
The proof is given in Appendix.
\end{IEEEproof}

According to item (ii) of Theorem \ref{Thm_4_1} and items (ii) and (iii) of Theorem \ref{Thm_4_2} (which indicate $\lambda_{i,0}\ne \lambda_{i,1}$ for any $m+1\le i\le 3m$), a finite field  $\mathbf{F}_q$ of odd characteristic
with at least $2m$ pairwise distinct nonzero elements is necessary
to ensure the code $\mathcal{C}_1$ to be an MSR code.
In the following theorem,  a class of concrete coefficients for  code $\mathcal{C}_1$ is given.

\begin{Theorem}\label{Thm_4_4}
The code $\mathcal{C}_1$ in Construction
\ref{Con_c4} is an MSR code if
\begin{eqnarray*}
&& k_i=k_{i+m}=-2\gamma^i, \ \lambda_{i,0}=\lambda_{i,1}=\lambda_{i+m,0}=\lambda_{i+2m,1}=\gamma^{i},\\
&& \lambda_{i+m,1}=\lambda_{i+2m,0}=-\gamma^{i},t_i=-1, t_{i+m}=1
\end{eqnarray*}
for  $1\le i\le m$,
where $\gamma$ is a primitive element of finite field $\mathbf{F}_q$ of odd characteristic with $q\ge 2m+1$. In particular, $q=\min\{p^i\ge 2m+1|p\ is\ an\ odd\ prime,\ i\ge1\}$ is the optimal alphabet size for $\mathcal{C}_1$ to be an MSR code.
\end{Theorem}
\begin{IEEEproof}
The proof is given in Appendix.
\end{IEEEproof}

\begin{Remark}
For a given storage capacity $\alpha=2^m$ per node,  our code $\mathcal{C}_1$ and  the long MDS code in \cite{Long}
have the biggest size $3m$ among all the MSR codes with high rate. Unlike the long MDS code,
$\mathcal{C}_1$ has $m$ systematic nodes  possessing the optimal access  property and the optimal update property simultaneously.
However, $\mathcal{C}_1$ may require a larger alphabet size than that of the long MDS code in certain situations
since only the finite field of odd characteristic is feasible for the construction of $\mathcal{C}_1$.
\end{Remark}

Finally, an illustrative example of code $\mathcal{C}_1$ is given.
\begin{Example}
For $m=2$, the coding matrices and repair matrices of the code $\mathcal{C}_1$ are as follows:
\begin{equation*}
    A_1=\left(
          \begin{array}{c}
            2e_2 \\
            2e_3 \\
            2e_0 \\
            2e_1 \\
          \end{array}
        \right),\ \ A_2=\left(
          \begin{array}{c}
            4e_1 \\
            4e_0 \\
            4e_3 \\
            4e_2 \\
          \end{array}
        \right),
\end{equation*}
\begin{equation*}
 A_3=\left(
          \begin{array}{c}
            2e_0 \\
            2e_1 \\
            e_0+3e_2 \\
            e_1+3e_3 \\
          \end{array}
        \right),\ \
  A_4=\left(
          \begin{array}{c}
            4e_0 \\
            2e_0+e_1 \\
            4e_2 \\
            2e_2+e_3 \\
          \end{array}
        \right),
 \end{equation*}
\begin{equation*}
A_5=\left(
          \begin{array}{c}
            3e_0 \\
            3e_1 \\
            e_0+2e_2 \\
            e_1+2e_3 \\
          \end{array}
        \right),\ \ A_6=\left(
          \begin{array}{c}
            e_0 \\
            2e_0+4e_1 \\
            e_2 \\
            2e_2+4e_3 \\
          \end{array}
        \right),
\end{equation*}
\begin{equation*}
    S_1=\left(
          \begin{array}{c}
            e_0 \\
            e_1 \\
          \end{array}
        \right),\ \ S_2=\left(
          \begin{array}{c}
            e_0 \\
            e_2 \\
          \end{array}
        \right),\ \ S_3=\left(
          \begin{array}{c}
            e_0-e_2 \\
            e_1-e_3 \\
          \end{array}
        \right),
\end{equation*}
\begin{equation*}
  S_4=\left(
          \begin{array}{c}
            e_0-e_1 \\
            e_2-e_3 \\
          \end{array}
        \right),\ \ S_5=\left(
          \begin{array}{c}
            e_0+e_2 \\
            e_1+e_3 \\
          \end{array}
        \right),\ \ S_6=\left(
          \begin{array}{c}
            e_0+e_1 \\
            e_2+e_3 \\
          \end{array}
        \right)
\end{equation*}
where  $2$ is chosen as a primitive element of $\mathbf{F}_{5}$ and all the calculations are done over $\mathbf{F}_{5}$. It can be easily verified that R1-R3 hold and R4-R5 hold for $1\le i\le m$, which are consistent with Theorems \ref{Thm_4_4} and \ref{Thm_4 3}, respectively.
\end{Example}

\subsection{New code $\mathcal{C}_2$}

Deleting the last $m$ systematic nodes in $\mathcal{C}_1$, we can get the second new code.

\begin{Construction}\label{Con_c3}
The $(n=k+2,k=2m)$ code $\mathcal{C}_2$ has $\alpha\times \alpha$ coding matrices $A_i$   and $\frac{\alpha}{2}\times \alpha$ repair matrices $S_i$ for $1\le i\le k$, such that
\begin{enumerate}
\item
$\ \ \ \left(
  \begin{array}{c}
    V_{i,0} \\
    V_{i,1} \\
  \end{array}
\right)A_i\\[4pt]=\left\{\begin{array}{ll}
\left(
             \begin{array}{c}
               \lambda_{i,1}V_{i,1} \\
               \lambda_{i,0}V_{i,0} \\
             \end{array}
           \right), &\mbox{if}\ \   1\le i\le m\\[12pt]
    \left(
                 \begin{array}{l}
                  \lambda_{i,0} V_{i,0} \\
                \lambda_{i,1}V_{i,1}+k_{i-m}V_{i,0}\\
                 \end{array}
               \right),&\mbox{if}\ \    m+1\le i\le 2m
\end{array}
\right.$
\item
$
 \ \ \    \left(
  \begin{array}{c}
    V_{i,0} \\
    V_{i,1} \\
  \end{array}
\right)A_i\\[4pt]=\left\{\begin{array}{ll}
\left(
             \begin{array}{c}
               \lambda_{i,1}V_{i,1} \\
               \lambda_{i,0}V_{i,0} \\
             \end{array}
           \right), &\mbox{if}\ \   1\le i\le m\\[12pt]
    \left(
                 \begin{array}{l}
                  \lambda_{i,0} V_{i,0} \\
                \lambda_{i,1}V_{i,1}+k_{i-m}V_{i,0}\\
                 \end{array}
               \right),&\mbox{if}\ \    m+1\le i\le 2m
\end{array}
\right.$
   \vspace{2mm}
\item
  $S_i=\left\{\begin{array}{ll}V_{i,0},&\mbox{if}\ \   1\le i\le m
\\
V_{i,0}+t_{i-m}V_{i,1},&\mbox{if}\ \  m+1\le i\le 2m \end{array}\right.$
   \vspace{2mm}
\end{enumerate}
where $\lambda_{i,0},\lambda_{i,1},k_j,t_j\in \mathbf{F}_q^*$ for all $1\le i\le k$ and $1\le j\le m$.
\end{Construction}

Hereafter we state
the results of $\mathcal{C}_2$ without proofs since they are included in those given in the last subsection.

\begin{Theorem}\label{Thm_3_1}
$\mathcal{C}_2$ is  a code with the MDS property if and only if
\begin{enumerate}
  \item [(i)] $\lambda_{i,0}\lambda_{i,1}\neq \lambda_{j,0}\lambda_{j,1}$ for any $1\le i\neq j\le m$,
  \item [(ii)] $\lambda_{i,s}\neq\lambda_{j,t}$ for any $m+1\le i\neq j\le 2m$ and $s,t=0,1$,
  \item [(iii)] $\left\{\begin{array}{ll}\lambda_{i,1}(\lambda_{i,0}-k_i)\neq \lambda_{j,0}\lambda_{j,1}, &\mbox{if}\ \ j=i+m \\ \lambda_{i,0}\lambda_{i,1}\neq\lambda_{j,0}^2,\lambda_{j,1}^2, &\mbox{if}\ \ j\neq i+m\end{array}\right.$\\   for any $1\le i\le m$ and  $m+1\le j\le 2m$.
\end{enumerate}
\end{Theorem}

\begin{Theorem}\label{Thm_3_2}
$\mathcal{C}_2$ is a code with the optimal repair property if and only if
\begin{enumerate}
  \item [(i)] $\lambda_{i,1}=t_i^2\lambda_{i,0}$ for all $1\le i\le m$,
  \item [(ii)] $\lambda_{i,1}\ne\lambda_{i,0}+t_{i-m}k_{i-m}$ for any $m+1\le i\le 2m$.
\end{enumerate}
\end{Theorem}

\begin{Theorem}\label{Thm_3-3}
The first $m$ systematic nodes of $\mathcal{C}_2$  have both the optimal access property and the optimal update property.
\end{Theorem}

According to item (i) of Theorem \ref{Thm_3_1} and item (i) of Theorem \ref{Thm_3_2}, a finite field  $\mathbf{F}_q$
with at least $m$ pairwise distinct nonzero square elements is necessary
to ensure the code $\mathcal{C}_2$ to be an MSR code. Let $q=p^i$ where $p$ is a prime
and $i$ is a positive integer. It is well known that all the nonzero elements in $\mathbf{F}_q$
are square elements for $p=2$ but only half the nonzero elements in $\mathbf{F}_q$
are square elements for $p>2$. Then, the MSR code $\mathcal{C}_2$
requires $q\ge m+1$ for $p=2$ or  $q\ge 2m+1$ for $p>2$. Straightforwardly,
there exits a positive integer $i$ such that $q=2^i$ lies between $m+1$ and  $2m$.
That is, a finite field of characteristic 2 is more suitable to construct the MSR code $\mathcal{C}_2$.
In the following theorem,  a class of concrete coefficients for  code $\mathcal{C}_2$ is given.

\begin{Theorem}\label{Thm_3_3}
The code $\mathcal{C}_2$ in Construction
\ref{Con_c3} is an MSR code if
\begin{equation*}
    \lambda_{i,0}=\lambda_{i,1}=\lambda_{i+m,0}=\lambda_{i+m,1}=\gamma^{i},\ \mbox{and}\  t_i=k_i=1
\end{equation*}
for all $ 1\le i\le m$,
where $\gamma$ is a primitive element of finite field $\mathbf{F}_q$ of characteristic 2 with $q\ge m+1$. In particular, $q=\min\{2^i\ge m+1|i\ge1\}$ is the optimal alphabet size for $\mathcal{C}_2$ to be an MSR code.
\end{Theorem}

An illustrative example of code $\mathcal{C}_2$ is given as follows.
\begin{Example}
For $m=3$, the coding matrices and repair matrices of the code $\mathcal{C}_2$ are as follows:
\begin{equation*}
A_1=\left(
          \begin{array}{c}
           \gamma e_4 \\
           \gamma e_5 \\
            \gamma e_6 \\
           \gamma e_7 \\
           \gamma e_0 \\
           \gamma  e_1 \\
          \gamma  e_2 \\
           \gamma e_3 \\
          \end{array}
        \right),\ \ A_2=\left(
          \begin{array}{c}
            \gamma^2 e_2 \\
            \gamma^2 e_3 \\
            \gamma^2e_0 \\
            \gamma^2e_1 \\
            \gamma^2 e_6 \\
            \gamma^2 e_7 \\
            \gamma^2e_4 \\
           \gamma^2 e_5 \\
          \end{array}
        \right),
\end{equation*}

\begin{equation*}
 A_3=\left(
          \begin{array}{c}
            e_1 \\
            e_0 \\
            e_3 \\
            e_2 \\
             e_5 \\
            e_4 \\
             e_7 \\
            e_6 \\
          \end{array}
        \right),\ \
A_4=\left(
          \begin{array}{l}
            \gamma e_0 \\
           \gamma  e_1 \\
           \gamma  e_2 \\
           \gamma  e_3 \\
           \gamma  e_4+e_0 \\
           \gamma  e_5+e_1 \\
           \gamma  e_6+e_2 \\
           \gamma  e_7+e_3 \\
          \end{array}
        \right),
\end{equation*}

\begin{equation*}
 A_5=\left(
          \begin{array}{l}
            \gamma^2 e_0 \\
            \gamma^2 e_1 \\
          \gamma^2 e_2+e_0 \\
            \gamma^2e_3+e_1 \\
            \gamma^2 e_4 \\
            \gamma^2 e_5 \\
           \gamma^2 e_6+e_4 \\
            \gamma^2 e_7+e_5 \\
          \end{array}
        \right),\ \ A_6=\left(
          \begin{array}{l}
            e_0 \\
            e_1+e_0 \\
           e_2 \\
            e_3+e_2 \\
            e_4 \\
           e_5+e_4 \\
            e_6 \\
            e_7+e_6 \\
          \end{array}
        \right),
\end{equation*}
\begin{equation*}
    S_1=\left(
          \begin{array}{c}
            e_0 \\
            e_1 \\
            e_2 \\
            e_3 \\
          \end{array}
        \right),\ \ S_2=\left(
          \begin{array}{c}
            e_0 \\
            e_1 \\
            e_4 \\
            e_5 \\
          \end{array}
        \right),
\end{equation*}
\begin{equation*}
 S_3=\left(
          \begin{array}{c}
            e_0 \\
            e_2 \\
            e_4 \\
            e_6 \\
          \end{array}
        \right),\ \
  S_4=\left(
          \begin{array}{c}
            e_0+e_4 \\
            e_1+e_5 \\
            e_2+e_6 \\
            e_3+e_7 \\
          \end{array}
        \right),
 \end{equation*}
\begin{equation*}
 S_5=\left(
          \begin{array}{c}
            e_0+e_2 \\
            e_1+e_3 \\
            e_4+e_6 \\
            e_5+e_7 \\
          \end{array}
        \right),\ \ S_6=\left(
          \begin{array}{c}
            e_0+e_1 \\
            e_2+e_3 \\
            e_4+e_5 \\
            e_6+e_7 \\
          \end{array}
        \right)
\end{equation*}
where $\gamma$ is chosen as a primitive element of $\mathbf{F}_{2^2}$ and all the calculations are done over $\mathbf{F}_{2^2}$. It can be verified that R1-R3 hold and R4-R5 hold for $1\le i\le m$, which are consistent with Theorems \ref{Thm_3_3} and \ref{Thm_3-3}, respectively.
\end{Example}

\subsection{New Code $\mathcal{C}_3$}

By means of combination of coding matrices of  types I and II, we propose the third new code.

\begin{Construction}\label{Con_2}
The $(n=k+2,k=2m)$ code $\mathcal{C}_3$ has $\alpha\times \alpha$ coding matrices $A_i$   and $\frac{\alpha}{2}\times \alpha$ repair matrices $S_i$ for $1\le i\le k$, such that
\begin{enumerate}
  \item $
    \left(
  \begin{array}{c}
    V_{i,0} \\
    V_{i,1} \\
  \end{array}
\right)A_i=\left\{\begin{array}{ll}
\left(
             \begin{array}{c}
               \lambda_{i,1}V_{i,1} \\
               \lambda_{i,0}V_{i,0} \\
             \end{array}
           \right), &\mbox{if}\ \  1\le i\le m\\[12pt]
    \left(
             \begin{array}{c}
               \lambda_{i,0}V_{i,0} \\
               \lambda_{i,1}V_{i,1} \\
             \end{array}
           \right),&\mbox{if}\ \   m+1\le i\le 2m
\end{array}
\right.$
\vspace{2mm}
  \item
  $S_i=\left\{\begin{array}{ll}V_{i,0},&\mbox{if\ \  $1\le i\le m$}
\\
V_{i,0}+t_{i-m}V_{i,1},&\mbox{if\ \  $m+1\le i\le 2m$} \end{array}\right.$
\end{enumerate} where $\lambda_{i,0},\lambda_{i,1},t_j\in \mathbf{F}_q^*$ for all $1\le i\le k$ and $1\le j\le m$.
\end{Construction}

\begin{Theorem}\label{Thm_1}
$\mathcal{C}_3$ is a code with the MDS property if and only if
\begin{enumerate}
  \item [(i)] $\lambda_{i,0}\lambda_{i,1}\neq \lambda_{j,0}\lambda_{j,1}$ for any $1\le i\neq j\le m$,
  \item [(ii)] $\lambda_{i,s}\neq\lambda_{j,t}$ for any  $m+1\le i\neq j\le 2m$ and $s,t=0$ or $1$,
  \item [(iii)] $\lambda_{i,0}\lambda_{i,1}\neq\left\{\begin{array}{ll}\lambda_{j,0}\lambda_{j,1}, &\mbox{if}\ \ j=i+m \\ \lambda_{j,0}^2, \lambda_{j,1}^2, &\mbox{if}\ \ j\neq i+m\end{array}\right.$\\   for any $1\le i\le m$ and  $m+1\le j\le 2m$.
\end{enumerate}
\end{Theorem}

\begin{IEEEproof}
The proof is given in Appendix.
\end{IEEEproof}

\begin{Theorem}\label{Thm_2}
$\mathcal{C}_3$ is a code with the optimal repair property if and only if
\begin{enumerate}
  \item [(i)] $\lambda_{i,1}=t_i^2\lambda_{i,0}$ for all $1\le i\le m$,
  \item [(ii)] $\lambda_{i,0}\neq \lambda_{i,1}$ for any $m+1\le i\le 2m$.
\end{enumerate}
 \end{Theorem}

\begin{IEEEproof}
The proof is given in Appendix.
\end{IEEEproof}

\begin{Theorem}\label{Thm_3}
The $2m$ systematic nodes of $\mathcal{C}_3$  have the optimal update property and the first $m$ nodes have the optimal access property.
\end{Theorem}

According to  item (ii) of Theorem \ref{Thm_1} and item (ii) of Theorem \ref{Thm_2}, a finite field $\mathbf{F}_q$
with at least $2m$ pairwise distinct nonzero elements is required
to guarantee  the code $\mathcal{C}_3$ to be an MSR code.  Specifically,
over $\mathbf{F}_q$ with $q\ge 2m+1$, we can give a class of concrete coefficients  for  code $\mathcal{C}_3$ as follows.

\begin{Theorem}\label{Thm_code1}
 The code $\mathcal{C}_3$ in Construction
\ref{Con_2} is an MSR code if
\begin{equation*}
    \lambda_{i,0}=\lambda_{i,1}=\lambda_{i+m,0}=\gamma^{i},\  \lambda_{i+m,1}=\gamma^{\lfloor\frac{q}{2}\rfloor+i},\ \mbox{and}\  t_i=1
\end{equation*}
for all $1\le i\le m$,
where $\gamma$ is a primitive element of ~$\mathbf{F}_q$ with $q\ge 2m+1$. In particular, $q=\min\{p^i\ge 2m+1|p\mbox{\,\,is\,\,a\,\,prime\,\,},i\ge1\}$ is the optimal alphabet size for $\mathcal{C}_3$ to be an MSR code.
\end{Theorem}
\begin{IEEEproof}
The proof is given in Appendix.
\end{IEEEproof}

Finally to illustrate the construction of code $\mathcal{C}_3$, we give an example.
\begin{Example}
For $m=2$, the coding matrices and repair matrices of the code $\mathcal{C}_3$ are as follows:
\begin{equation*}
    A_1=\left(
          \begin{array}{c}
            2e_2 \\
            2e_3 \\
            2e_0 \\
            2e_1 \\
          \end{array}
        \right),\ \ A_2=\left(
          \begin{array}{c}
            4e_1 \\
            4e_0 \\
            4e_3 \\
            4e_2 \\
          \end{array}
        \right),
\end{equation*}
\begin{equation*}
 A_3=\left(
          \begin{array}{c}
            2e_0 \\
            2e_1 \\
            3e_2 \\
            3e_3 \\
          \end{array}
        \right),\ \ A_4=\left(
          \begin{array}{c}
            4e_0 \\
            e_1 \\
            4e_2 \\
            e_3 \\
          \end{array}
        \right),
\end{equation*}
\begin{equation*}
    S_1=\left(
          \begin{array}{c}
            e_0 \\
            e_1 \\
          \end{array}
        \right),\ \ S_2=\left(
          \begin{array}{c}
            e_0 \\
            e_2 \\
          \end{array}
        \right),
\end{equation*}
\begin{equation*}
 S_3=\left(
          \begin{array}{c}
            e_0+e_2 \\
            e_1+e_3 \\
          \end{array}
        \right),\ \ S_4=\left(
          \begin{array}{c}
            e_0+e_1 \\
            e_2+e_3 \\
          \end{array}
        \right),
\end{equation*}
where  $2$ is chosen as a primitive element of $\mathbf{F}_{5}$ and all the calculations are done over $\mathbf{F}_{5}$. It can be easily verified that R1-R3 hold and R4 holds for $1\leq i\leq m$ and R5 holds for $1\leq i\leq 2m$, which are consistent with Theorems \ref{Thm_code1} and \ref{Thm_3}, respectively. Moreover, this example can be illustrated in another way as in Table IV.

\begin{table}[htbp]\label{table update}
\centering
\caption{Columns $1,2,3,4$ are systematic nodes and columns R and Z are parity nodes. Each element in column R is a linear combination of the systematic elements in the same row, while each element in column Z is a linear combination of the systematic elements with the same symbol. For instance, the first element in column R is a linear combination of the elements in the first row and in columns 1,2,3 and 4, and the $\clubsuit$ in column Z is a linear combination of all the $\clubsuit$ elements in columns 1,2,3 and 4.}
\begin{tabular}{|c|c|c|c|c|c|c|}
\hline  & 1 & 2 & 3 &4 & R & Z \\
\hline 0& $\spadesuit$ & $\heartsuit$ & $\clubsuit$ &$\clubsuit$ &  &$\clubsuit$ \\
\hline  1& $\diamondsuit$ & $\clubsuit$ &$\heartsuit$ &$\heartsuit$ &  &  $\heartsuit$\\
\hline  2& $\clubsuit$ & $\diamondsuit$ & $\spadesuit$ &$\spadesuit$ &  & $\spadesuit$ \\
\hline  3& $\heartsuit$ &$\spadesuit$  & $\diamondsuit$ &$\diamondsuit$ &  &$\diamondsuit$  \\
\hline
\end{tabular}
\end{table}
\end{Example}

\subsection{New code $\mathcal{C}_4$}

Based on the coding matrices of  type II, we can present the fourth new code.

\begin{Construction}\label{Con_c2}
The $(n=k+2,k=2m)$ code $\mathcal{C}_4$ has $\alpha\times \alpha$ coding matrices $A_i$  and $\frac{\alpha}{2}\times \alpha$ repair matrices $S_i$  for $1\le i\le k$, such that
\begin{enumerate}
  \item $
    \left(
  \begin{array}{c}
    V_{i,0} \\
    V_{i,1} \\
  \end{array}
\right)A_i=\left(
             \begin{array}{c}
               \lambda_{i,1}V_{i,1} \\
               \lambda_{i,0}V_{i,0} \\
             \end{array}
           \right)$ for $1\le i\le k$,
\vspace{2mm}
\item
  $S_i=V_{i,0}+t_{i}V_{i,1}$ for $1\le i\le k$,
\end{enumerate} where $\lambda_{i,0},\lambda_{i,1},t_i\in \mathbf{F}_q^*$ for all $1\le i\le k$.
\end{Construction}

\begin{Theorem}\label{Thm_2_1}
$\mathcal{C}_4$ is  a code with the MDS property if and only if
\begin{enumerate}
  \item [(i)] $\lambda_{i,0}\lambda_{i,1}\neq \lambda_{j,0}\lambda_{j,1}$ for any  $1\le i<j\le k$ and $i\neq j-m$,
  \item [(ii)] $\lambda_{i,s}\neq \lambda_{i+m,s}$ for any $1\le i\le m$ and $s=0,1$.
\end{enumerate}
\end{Theorem}

\begin{IEEEproof}
The proof is given in Appendix.
\end{IEEEproof}

\begin{Theorem}\label{Thm_2_2}
$\mathcal{C}_4$ is a code with the optimal repair property if and only if
\begin{enumerate}
  \item [(i)] $\lambda_{i,1}=t_{i+m}^2\lambda_{i,0}$ and $\lambda_{i+m,1}=t_i^2\lambda_{i+m,0}$ for all $1\le i\le m$,
  \item [(ii)] $\lambda_{i,1}\neq t_i^2\lambda_{i,0}$ for any $1\le i\le k$.
\end{enumerate}
 \end{Theorem}

\begin{IEEEproof}
The proof is given in Appendix.
\end{IEEEproof}

\begin{Theorem}\label{Thm_2-3}
The $2m$ systematic nodes of $\mathcal{C}_4$  have the optimal update property.
\end{Theorem}

According to item (i) of Theorem \ref{Thm_2_1} and item (i) of Theorem \ref{Thm_2_2}, a finite field $\mathbf{F}_q$
with at least $m$ pairwise distinct nonzero square elements is necessary
to ensure the code $\mathcal{C}_4$ to be an MSR code.  Similar to
code $\mathcal{C}_2$, we have the  following concrete construction for the new code $\mathcal{C}_4$.

\begin{Theorem}\label{Thm_2_3}
The code $\mathcal{C}_4$ in Construction
\ref{Con_c2} is an MSR code if
{\xiaowuhao  \begin{equation*}
  \lambda_{i,0}=\gamma^i,\lambda_{i,1}=\gamma^{i+2}, \lambda_{i+m,0}=\lambda_{i+m,1}=\gamma^{i+1}, t_i=1, t_{i+m}=\gamma
\end{equation*}
}
for all $1\le i\le m$, where $\gamma$ is a primitive element of finite field $\mathbf{F}_q$ of characteristic 2  with $q\ge m+1$. In particular, $q=\min\{2^i\ge m+1|i\ge1\}$ is the optimal alphabet size for $\mathcal{C}_4$ to be an MSR code.
\end{Theorem}

\begin{Remark}
In \cite{Zigzag}, 2-duplication of the Zigzag code with parameters $(n=k+2,k=2m+2)$ was proposed, which has the similar coding matrices as those of $\mathcal{C}_4$. Although the number of systematic nodes of $\mathcal{C}_4$ is two less than that of 2-duplication of the Zigzag code, it has better repair bandwidth.  When repairing a failed systematic node, only half of the data need to be downloaded from each surviving node of $\mathcal{C}_4$, while the fraction of the data need to be downloaded from each surviving node of  2-duplication of the Zigzag code is $\frac{m+1}{2m+1}$.
\end{Remark}

Finally to illustrate the construction of code $\mathcal{C}_4$, we give an example.
\begin{Example}
For $m=2$, the coding matrices and repair matrices of the code $\mathcal{C}_4$ are as follows:
\begin{equation*}
    A_1=\left(
          \begin{array}{c}
            e_2 \\
            e_3 \\
            \gamma e_0 \\
            \gamma e_1 \\
          \end{array}
        \right),\ \ A_2=\left(
          \begin{array}{c}
            \gamma e_1 \\
            \gamma^2 e_0 \\
            \gamma e_3 \\
            \gamma^2 e_2 \\
          \end{array}
        \right),
\end{equation*}
\begin{equation*} A_3=\left(
          \begin{array}{c}
            \gamma^2 e_2 \\
            \gamma^2 e_3 \\
            \gamma^2 e_0 \\
            \gamma^2 e_1 \\
          \end{array}
        \right),\ \ A_4=\left(
          \begin{array}{c}
            e_1 \\
            e_0 \\
            e_3 \\
            e_2 \\
          \end{array}
        \right),
\end{equation*}
\begin{equation*}
    S_1=\left(
          \begin{array}{c}
            e_0+e_2 \\
            e_1+e_3 \\
          \end{array}
        \right),\ \ S_2=\left(
          \begin{array}{c}
            e_0+e_1 \\
            e_2+e_3 \\
          \end{array}
        \right),
\end{equation*}
\begin{equation*}
 S_3=\left(
          \begin{array}{c}
            e_0+\gamma e_2 \\
            e_1+\gamma e_3 \\
          \end{array}
        \right),\ \ S_4=\left(
          \begin{array}{c}
            e_0+\gamma e_1 \\
            e_2+\gamma e_3 \\
          \end{array}
        \right)
\end{equation*}
where $\gamma$ is chosen as a primitive element of $\mathbf{F}_{2^2}$ and all the calculations are done over $\mathbf{F}_{2^2}$. It can be verified that R1-R3 hold and R5 holds for $1\le i\le 2m$, which are consistent with Theorems \ref{Thm_2_3} and \ref{Thm_2-3}, respectively.
Moreover, this example can be illustrated in another way as in the following table.
\begin{center}
\begin{tabular}{|c|c|c|c|c|c|c|}
\hline  & 1 & 2 & 3 &4 & R & Z \\
\hline 0& $\spadesuit$ & $\heartsuit$ & $\spadesuit$ &$\heartsuit$ &  &$\clubsuit$ \\
\hline  1& $\diamondsuit$ & $\clubsuit$ &$\diamondsuit$ &$\clubsuit$ &  &  $\heartsuit$\\
\hline  2& $\clubsuit$ & $\diamondsuit$ & $\clubsuit$ &$\diamondsuit$ &  & $\spadesuit$ \\
\hline  3& $\heartsuit$ &$\spadesuit$  & $\heartsuit$ &$\spadesuit$ &  &$\diamondsuit$  \\
\hline
\end{tabular}
\end{center}
\end{Example}

\subsection{Other new codes}

Combined coding matrices of types I and IV (with repair matrices $V_{i,0}+t_iV_{i,1}$ and $V_{i,0}$), types II and IV (with repair matrices $V_{i,0}+t_iV_{i,1}$), types III and IV (with repair matrices $V_{i,0}+t_iV_{i,1}$ and $V_{i,0}$), types III, III and IV (with repair matrices $V_{i,0}+t_iV_{i,1}$, $V_{i,0}+t_{i+m}V_{i,1}$ and $V_{i,0}$),  four new MSR codes with $k=2m$ or $3m$ can be obtained, but the other properties (eg. optimal access, optimal update, the size of the finite fields required) are not as good as the aforementioned new codes $\mathcal{C}_1$, $\mathcal{C}_2$, $\mathcal{C}_3$ and $\mathcal{C}_4$.

\section{Concluding remarks}

In this paper, we proposed a simple but generic  framework to construct high rate MSR codes with two parity nodes. The framework can not only generate the modified Zigzag code  and the long MDS code, but also generate four new MSR codes $\mathcal{C}_1$, $\mathcal{C}_2$ , $\mathcal{C}_3$ and $\mathcal{C}_4$ with the optimal access/update property. The optimal sizes of the finite fields required for the four codes were also determined. Notably, by these four
new MSR codes, we could get a tradeoff  between
the size of the finite field and the number of systematic
nodes (with the optimal access/update property).

 Our construction  can be generalized to the $(k+r,k=3m)$ or $(k+r,k=2m)$ MSR code with arbitrary $r> 2$ parity nodes for $\alpha=r^m$. For this generalization, we firstly need to partition the basis $\{e_0,e_1,\cdots,e_{r^m-1}\}$ of $\mathbf{F}_q^\alpha$ into $r$ subsets $V_0,V_1,\cdots,V_{r}$ with equal sizes. Then, types I-IV coding matrices can be similarly determined based on invariant subspaces of dimension $r$ but with complicated forms. By means of these matrices, we can  obtain the
generalized codes $\mathcal{C}_1$, $\mathcal{C}_2$, $\mathcal{C}_3$ and $\mathcal{C}_4$ with $r>2$ parity nodes, which
still possesses the optimal access/update property.
The optimal alphabet size $q$, however, is difficult to determine and hence will be left for future research.

\section*{Appendix}

\textbf{Proof of Lemma 1}:
(i) According to \eqref{Eqn_B3}, in matrix notation, $V_{i,0},V_{i,1}$ are equivalent to  $\left(
           \begin{array}{c}
             V_{i,j,0,0} \\
             V_{i,j,0,1} \\
           \end{array}
         \right)
$ and $\left(
           \begin{array}{c}
             V_{i,j,1,0} \\
             V_{i,j,1,1} \\
           \end{array}
         \right)
$ under elementary row transformation, respectively, i.e., $\left(
                                                              \begin{array}{c}
                                                                V_{i,0} \\
                                                                V_{i,1} \\
                                                              \end{array}
                                                            \right)
$ is  equivalent to $\left(
           \begin{array}{c}
             V_{i,j,0,0} \\
             V_{i,j,0,1} \\
              V_{i,j,1,0} \\
             V_{i,j,1,1} \\
           \end{array}
         \right)$ under elementary row transformation. Thus
\begin{equation*}
    \mbox{rank}\left(\left(
                                                              \begin{array}{c}
                                                                V_{i,0} \\
                                                                V_{i,1} \\
                                                              \end{array}
                                                            \right)\right)=\mbox{rank}\left(\left(
           \begin{array}{c}
             V_{i,j,0,0} \\
             V_{i,j,0,1} \\
              V_{i,j,1,0} \\
             V_{i,j,1,1} \\
           \end{array}
         \right)\right).
\end{equation*}
Immediately, the assertion follows from  the fact that the matrix $\left(
                                                              \begin{array}{c}
                                                                V_{i,0} \\
                                                                V_{i,1} \\
                                                              \end{array}
                                                            \right)$  is of full rank.

(ii) It follows from that the matrix $\left(
                         \begin{array}{l}
                           V_{i,0}+u_iV_{i,1} \\
                           (V_{i,0}+u_iV_{i,1})A_j \\
                         \end{array}
                       \right)$ is equivalent to $\left(
                         \begin{array}{l}
                           V_{i,j,0,0}+u_iV_{i,j,1,0} \\
                           V_{i,j,0,1}+u_iV_{i,j,1,1} \\
                           (V_{i,j,0,0}+u_iV_{i,j,1,0})A_j \\
                           (V_{i,j,0,1}+u_iV_{i,j,1,1})A_j \\
                         \end{array}\right)$
under elementary row transformation.
\hfill$\blacksquare$

\textbf{Proof of Proposition 1}: (i) It is obvious otherwise R2 and R3 can not be satisfied simultaneously for repair matrices
$S_i, S_j$ and coding matrix $A_i$.

(ii) If there exist such four repair matrices, according to the generic construction,  then  the coding matrix $A_{j_4}$ satisfies
\begin{eqnarray*}
    \left(
  \begin{array}{c}
    V_{i,0} \\
    V_{i,1} \\
  \end{array}
\right)A_{j_4}=\left(
             \begin{array}{c}
               a V_{i,0}+b V_{i,1} \\
               c V_{i,0}+d V_{i,1} \\
             \end{array}
           \right)
\end{eqnarray*}
where $a$, $b$, $c$ and $d$ can be coefficients in $\mathbf{F}_q$ or $\frac{\alpha}{2}\times \frac{\alpha}{2}$ diagonal matrices  over $\mathbf{F}_q$. Consider
\begin{eqnarray*}
  && \mbox{rank} \left(
  \begin{array}{c}
    V_{i,0}+t_l V_{i,1} \\
    (V_{i,0}+t_l V_{i,1})A_{j_4} \\
  \end{array}
\right)\\&=&\mbox{rank} \left(
             \begin{array}{c}
               V_{i,0}+t_l V_{i,1}\\
               (a+c t_l) V_{i,0}+ ( b+d t_l) V_{i,1} \\
             \end{array}
           \right)\\&=&\frac{\alpha}{2}, ~l=1,2,3.
\end{eqnarray*}
Then we have that the equation
\begin{equation*}
  ct^2+(a-d)t-b=0
\end{equation*}
has three distinct roots $t=t_1,t_2$ and $t_3$,  which is possible only if
$b=c=0$ and $a=d\ne 0$, in this case, $A_{j_4}$ is a diagonal matrix, therefore
\begin{eqnarray*}
     \mbox{rank} \left(
  \begin{array}{c}
   S_{j_4} \\
    S_{j_4}A_{j_4} \\
  \end{array}
\right)={\alpha\over 2}\ \ \mbox{for\ any\ }S_{j_4}
\end{eqnarray*}
and then R3 can not be satisfied.
\hfill$\blacksquare$

\textbf{Proof of Theorem \ref{Thm_4_1}}:
$\mathcal{C}_1$ has the MDS property if and only if R1 holds. Obviously, $A_i$ is invertible for all $1\le i\le k$ since $\lambda_{i,0},\lambda_{i,1}\ne0$. In what follows, by means of  Lemma \ref{rank} we establish  the necessary and sufficient conditions of $\mbox{rank}(A_i-A_j)=\alpha$ for any $1\le i\ne j\le k$ in the following three cases.

Case 1: When $1\le i\neq j\le m$,
  \begin{equation*}
    \left(
       \begin{array}{c}
         V_{i,j,0,0} \\
V_{i,j,0,1} \\
V_{i,j,1,0} \\
V_{i,j,1,1} \\
       \end{array}
     \right)(A_i-A_j)=\left(
                   \begin{array}{l}
                     \lambda_{i,1}V_{i,j,1,0}-\lambda_{j,1}V_{i,j,0,1} \\
                     \lambda_{i,1}V_{i,j,1,1}-\lambda_{j,0}V_{i,j,0,0} \\
                     \lambda_{i,0}V_{i,j,0,0}-\lambda_{j,1}V_{i,j,1,1} \\
                     \lambda_{i,0}V_{i,j,0,1}-\lambda_{j,0}V_{i,j,1,0}\\
                   \end{array}
                 \right)
\end{equation*}
i.e., $
    \mbox{rank}(A_i-A_j)=\alpha\Leftrightarrow\lambda_{i,0}\lambda_{i,1}\ne\lambda_{j,0}\lambda_{j,0}$.
\vspace{2mm}

Case 2: When $m+1\le i< j\le 3m$, if $j=i+m$, by (\ref{B3})
\begin{eqnarray*}
  &&\left(
       \begin{array}{c}
         V_{i,0} \\
         V_{i,1} \\
       \end{array}
     \right)(A_i-A_j)\\ &=& \left(
       \begin{array}{l}
         \lambda_{i,0}V_{i,0} \\
\lambda_{i,1}V_{i,1}+k_{i-m}V_{i,0} \\
       \end{array}
     \right)-\left(
       \begin{array}{l}
         \lambda_{j,0}V_{i,0} \\
\lambda_{j,1}V_{i,1}+k_{j-m}V_{i,0} \\
       \end{array}
     \right) \\
  &=&\left(
                          \begin{array}{l}
                           (\lambda_{i,0}-\lambda_{j,0})V_{i,0} \\
                            (\lambda_{i,1}-\lambda_{j,1})V_{i,1}+(k_{i-m}-k_{j-m})V_{i,0}\\
                          \end{array}
                        \right)
\end{eqnarray*}
i.e., $\mbox{rank}(A_i-A_j)=\alpha\Leftrightarrow\lambda_{i,s}\ne \lambda_{j,s}\ \mbox{for\ }s=0,1$;
Otherwise,
  \begin{eqnarray*}
    &&\hspace{-2mm}\left(
       \begin{array}{c}
         V_{i,j,0,0} \\
V_{i,j,0,1} \\
V_{i,j,1,0} \\
V_{i,j,1,1} \\
       \end{array}
     \right)(A_i-A_j)\\\hspace{-2mm}&=&\hspace{-2mm}\left(
                   \begin{array}{l}
                     (\lambda_{i,0}-\lambda_{j,0})V_{i,j,0,0} \\
                     (\lambda_{i,0}-\lambda_{j,1})V_{i,j,0,1}-k_{j-m}V_{i,j,0,0} \\
                     (\lambda_{i,1}-\lambda_{j,0})V_{i,j,1,0}+k_{i-m}V_{i,j,0,0} \\
                     (\lambda_{i,1} -\lambda_{j,1})V_{i,j,1,1}+k_{i-m}V_{i,j,0,1}-k_{j-m}V_{i,j,1,0}\\
                   \end{array}
                 \right)
\end{eqnarray*}
i.e., $
    \mbox{rank}(A_i-A_j)=\alpha\Leftrightarrow\lambda_{i,s}\ne\lambda_{j,t}\ \mbox{for\ }s,t=0,1$.
\vspace{2mm}

Case 3: When  $1\le i\le m$ and  $m+1\le j\le 3m$, if $j=i+m$ or $i+2m$, according to (\ref{B3})
\begin{eqnarray*}
  &&\left(
       \begin{array}{c}
         V_{i,0} \\
         V_{i,1} \\
       \end{array}
     \right)(A_i-A_j)\\&=& \left(
       \begin{array}{l}
         \lambda_{i,1}V_{i,1} \\
\lambda_{i,0}V_{i,0}\\
       \end{array}
     \right)-\left(
       \begin{array}{l}
         \lambda_{j,0}V_{i,0} \\
\lambda_{j,1}V_{i,1}+k_{j-m}V_{i,0} \\
       \end{array}
     \right) \\
  &=&\left(
                          \begin{array}{l}
                           \lambda_{i,1}V_{i,1}-\lambda_{j,0}V_{i,0} \\
                            (\lambda_{i,0}-k_{j-m})V_{i,0}-\lambda_{j,1}V_{i,1}\\
                          \end{array}
                        \right),
\end{eqnarray*}
i.e., $
    \mbox{rank}(A_i-A_j)=\alpha\Leftrightarrow\lambda_{i,1}(\lambda_{i,0}-k_{j-m})\neq \lambda_{j,0}\lambda_{j,1}$;
Otherwise,
  \begin{eqnarray*}
    &&\left(
       \begin{array}{c}
         V_{i,j,0,0} \\
V_{i,j,0,1} \\
V_{i,j,1,0} \\
V_{i,j,1,1} \\
       \end{array}
     \right)(A_i-A_j)\\&=&\left(
                   \begin{array}{l}
\lambda_{i,1}V_{i,j,1,0}-\lambda_{j,0}V_{i,j,0,0} \\
\lambda_{i,1}V_{i,j,1,1}-\lambda_{j,1}V_{i,j,0,1}-k_{j-m}V_{i,j,0,0} \\
\lambda_{i,0}V_{i,j,0,0}-\lambda_{j,0}V_{i,j,1,0} \\
\lambda_{i,0}V_{i,j,0,1} -\lambda_{j,1}V_{i,j,1,1}-k_{j-m}V_{i,j,1,0}\\
                   \end{array}
                 \right)
\end{eqnarray*}
i.e., $
    \mbox{rank}(A_i-A_j)=\alpha\Leftrightarrow\lambda_{i,0}\lambda_{i,1}\neq\lambda_{j,0}^2,\lambda_{j,1}^2$.
\hfill$\blacksquare$

\textbf{Proof of Theorem \ref{Thm_4_2}}:
$\mathcal{C}_1$ is a code with the optimal repair  property if and only if R2 and R3 hold.
Firstly, by means of  Lemma \ref{rank} we establish the necessary and sufficient conditions for
R2 according to the following three cases.

Case 1: For $1\le i\le m$,
  \begin{enumerate}
    \item [(a)] When $1\le j\ne i \le m$,
    \begin{eqnarray*}
     &&\mbox{rank}\left(\left(
                 \begin{array}{c}
                   S_i \\
                   S_iA_j \\
                 \end{array}
               \right)\right)\\&=&\mbox{rank}\left(\left(
                             \begin{array}{l}
                              V_{i,j,0,0} \\
                               V_{i,j,0,1} \\
                              V_{i,j,0,0}A_j \\
                               V_{i,j,0,1}A_j \\
                             \end{array}
                           \right)\right)\\
                           &=&\mbox{rank}\left(\left(
                             \begin{array}{l}
                              V_{i,j,0,0} \\
                               V_{i,j,0,1} \\
                             \lambda_{j,1} V_{i,j,0,1}\\
                             \lambda_{j,0}  V_{i,j,0,0} \\
                             \end{array}
                           \right)\right)\\
&=&\alpha/2,
\end{eqnarray*}
    \item [(b)] When $m+1\le j \le 3m$, if $j=i+m$ or $i+2m$, by (\ref{B3})
    \begin{eqnarray*}
     &&\mbox{rank}\left(\left(
                 \begin{array}{c}
                   S_i \\
                   S_iA_j \\
                 \end{array}
               \right)\right)\\&=&\mbox{rank}\left(\left(
                             \begin{array}{c}
                              V_{i,0}\\
                               V_{i,0}A_j \\
                             \end{array}
                           \right)\right)\\
                           &=&\mbox{rank}\left(\left(
                             \begin{array}{c}
                              V_{j,0}\\
                              \lambda_{j,0} V_{j,0} \\
                             \end{array}
                           \right)\right)\\
                           &=&\alpha/2;
\end{eqnarray*}
Otherwise,
\begin{eqnarray*}
     &&\mbox{rank}\left(\left(
                 \begin{array}{c}
                   S_i \\
                   S_iA_j \\
                 \end{array}
               \right)\right)\\&=&\mbox{rank}\left(\left(
                             \begin{array}{l}
                              V_{i,j,0,0} \\
                               V_{i,j,0,1} \\
                              V_{i,j,0,0}A_j \\
                               V_{i,j,0,1}A_j \\
                             \end{array}
                           \right)\right)\\
                           &=&\mbox{rank}\left(\left(
                             \begin{array}{l}
                              V_{i,j,0,0} \\
                               V_{i,j,0,1} \\
                              \lambda_{j,0}V_{i,j,0,0} \\
                               \lambda_{j,1}V_{i,j,0,1}+k_{j-m}V_{i,j,0,0}\\
                             \end{array}
                           \right)\right)\\
                           &=&\alpha/2.
\end{eqnarray*}
\end{enumerate}
\vspace{2mm}

Case 2: For $m+1\le i\le 2m$,
\begin{enumerate}
  \item [(a)] When $1\le j\le m$, if $j=i-m$, by (\ref{B3})
\begin{equation*}
    \begin{array}{rcl}
 &&\mbox{rank}\left(\left(
                 \begin{array}{c}
                   S_i \\
                   S_iA_j \\
                 \end{array}
               \right)\right)\\[12pt]&=&\mbox{rank}\left(\left(
                             \begin{array}{l}
                               V_{j,0}+t_{j}V_{j,1} \\
                               (V_{j,0}+t_{j}V_{j,1})A_j \\
                             \end{array}
                           \right)\right)\\[12pt]&=&\mbox{rank}\left(\left(
                                                     \begin{array}{l}
                                                      V_{j,0}+t_{j}V_{j,1} \\
                                                     \lambda_{j,1}V_{j,1}+t_{j}\lambda_{j,0}V_{j,0}
                                                     \end{array}
                                                   \right)\right)\\
&=&\alpha/2\\
&\Leftrightarrow&\lambda_{j,1}=t_j^2\lambda_{j,0};
\end{array}
\end{equation*}
Otherwise,
\begin{equation*}
    \begin{array}{rcl}
 &&\mbox{rank}\left(\left(
                 \begin{array}{c}
                   S_i \\
                   S_iA_j \\
                 \end{array}
               \right)\right)\\&=&\mbox{rank}\left(\left(
                             \begin{array}{l}
                              V_{i,j,0,0}+t_{i-m}V_{i,j,1,0} \\
                              V_{i,j,0,1}+t_{i-m}V_{i,j,1,1} \\
                               (V_{i,j,0,0}+t_{i-m}V_{i,j,1,0})A_j \\
                              (V_{i,j,0,1}+t_{i-m}V_{i,j,1,1})A_j \\
                             \end{array}
                           \right)\right)\\[20pt]
                           &=&\mbox{rank}\left(\left(
                             \begin{array}{l}
                              V_{i,j,0,0}+t_{i-m}V_{i,j,1,0} \\
                              V_{i,j,0,1}+t_{i-m}V_{i,j,1,1} \\
                               \lambda_{j,1}(V_{i,j,0,1}+t_{i-m}V_{i,j,1,1})\\
                              \lambda_{j,0}(V_{i,j,0,0}+t_{i-m}V_{i,j,1,0})\\
                             \end{array}
                           \right)\right)\\
&=&\alpha/2.
\end{array}
\end{equation*}
  \item [(b)] When $m+1\le j\ne i\le 3m$, if $j=i+m$, by (\ref{B3})
\xiaowuhao \begin{eqnarray*}
&&\hspace{-2mm} \mbox{rank}\left(\left(
                 \begin{array}{c}
                   S_i \\
                   S_iA_j \\
                 \end{array}
               \right)\right)\\[12pt]\hspace{-2mm}&=&\hspace{-2mm}\mbox{rank}\left(\left(
                             \begin{array}{l}
                               V_{j,0}+t_{i-m}V_{j,1} \\
                               (V_{j,0}+t_{i-m}V_{j,1})A_j \\
                             \end{array}
                           \right)\right)\\[12pt]\hspace{-2mm}&=&\hspace{-2mm}\mbox{rank}\left(\left(
                                                     \begin{array}{l}
                                                      V_{j,0}+t_{i-m}V_{j,1} \\
                                                     (\lambda_{j,0}+t_{i-m}k_{j-m})V_{j,0}+t_{i-m}\lambda_{j,1}V_{j,1}
                                                     \end{array}
                                                   \right)\right)\\
\hspace{-2mm}&=&\hspace{-2mm}\alpha/2\\
\hspace{-2mm}&\Leftrightarrow&\hspace{-2mm}\lambda_{j,1}=\lambda_{j,0}+t_{i-m}k_{j-m}\\
\hspace{-2mm}&\Leftrightarrow&\hspace{-2mm}\lambda_{i+m,1}=\lambda_{i+m,0}+t_{i-m}k_{i};
\end{eqnarray*}
Otherwise, \begin{eqnarray*}
&& \mbox{rank}\left(\left(
                 \begin{array}{c}
                   S_i \\
                   S_iA_j \\
                 \end{array}
               \right)\right)\\\hspace{-2mm}&=&\hspace{-2mm}\mbox{rank}\left(\left(
                             \begin{array}{l}
                              V_{i,j,0,0}+t_{i-m}V_{i,j,1,0} \\
                              V_{i,j,0,1}+t_{i-m}V_{i,j,1,1} \\
                               (V_{i,j,0,0}+t_{i-m}V_{i,j,1,0})A_j \\
                              (V_{i,j,0,1}+t_{i-m}V_{i,j,1,1})A_j \\
                             \end{array}
                           \right)\right)\\\hspace{-2mm}&=&\hspace{-2mm}\mbox{rank}\left(\left(
                             \begin{array}{l}
                              V_{i,j,0,0}+t_{i-m}V_{i,j,1,0} \\
                              V_{i,j,0,1}+t_{i-m}V_{i,j,1,1} \\
                               \lambda_{j,0}(V_{i,j,0,0}+t_{i-m}V_{i,j,1,0}) \\
                             \lambda_{j,1} (V_{i,j,0,1}+t_{i-m}V_{i,j,1,1})\\
                             \hspace{1cm}+k_{j-m}(V_{i,j,0,0}+t_{i-m}V_{i,j,1,0}) \\
                             \end{array}
                           \right)\right)\\
\hspace{-2mm}&=&\hspace{-2mm}\alpha/2.
\end{eqnarray*}
\end{enumerate}
\vspace{2mm}

Case 3: For $2m+1\le i\le 3m$, similarly to that of Case 2,
\begin{eqnarray*}
    && \mbox{rank}\left(\left(
                 \begin{array}{c}
                   S_i \\
                   S_iA_j \\
                 \end{array}
               \right)\right)=\alpha/2\,\,\mbox{for}\,\,1\le j\ne i\le 3m\\
&\Leftrightarrow&\lambda_{l,1}=\left\{\begin{array}{ll}t_{l+m}^2\lambda_{l,0},\ \mbox{if\ } 1\le l\le m,
\\
\lambda_{l,0}+t_{l}k_{l-m},\ \mbox{if\ } m+1\le l\le 2m. \end{array}\right.
\end{eqnarray*}

Combing all the cases above, we have that R2 holds if and only if
\begin{equation*}
  \lambda_{i,1}=t_i^2\lambda_{i,0}\ \mbox{for}\ 1\le i\le m,
\end{equation*}
\begin{equation}\label{G1}
  t_i^2=t_{i+m}^2\ \mbox{for}\ 1\le i\le m,
\end{equation}
and
\begin{equation}\label{G2}
  \lambda_{i,1}=\lambda_{i,0}+t_{i}k_{i-m},\ \lambda_{i+m,1}=\lambda_{i+m,0}+t_{i-m}k_{i}
\end{equation}
for $m+1\le i\le 2m$.

Secondly, we determine the necessary and sufficient conditions for
R3.
It is easy to verify that $\mbox{rank}\left(\left(
                 \begin{array}{c}
                   S_i \\
                   S_iA_i \\
                 \end{array}
               \right)\right)=\alpha$ for $1\le i\le m$.
For $m+1\le i\le 3m$, \begin{eqnarray*}
 &&\mbox{rank}\left(\left(
                 \begin{array}{c}
                   S_i \\
                   S_iA_i \\
                 \end{array}
               \right)\right)\\&=&\mbox{rank}\left(\left(
                             \begin{array}{l}
                               V_{i,0}+t_{i-m}V_{i,1} \\
                               (V_{i,0}+t_{i-m}V_{i,1})A_i \\
                             \end{array}
                           \right)\right)\\[12pt]&=&\mbox{rank}\left(\left(
                                                     \begin{array}{l}
                                                      V_{i,0}+t_{i-m}V_{i,1} \\
                                                       (\lambda_{i,0}+t_{i-m}k_{i-m})V_{i,0}+t_{i-m}\lambda_{i,1}V_{i,1}
                                                     \end{array}
                                                   \right)\right)\\
&=&\alpha\\
&\Leftrightarrow&\lambda_{i,1}\ne \lambda_{i,0}+t_{i-m}k_{i-m},
\end{eqnarray*}
which together with (\ref{G2}) gives $t_j\ne t_{j+m}$ for any $1\le j\le m$, and further, associated with (\ref{G1}) implies that $t_j=-t_{j+m}$ for all $1\le j\le m$ and $\mathbf{F}_q$ should be a finite field of odd characteristic. This finishes the proof.
\hfill$\blacksquare$

\textbf{Proof of Theorem \ref{Thm_4 3}}:
It is easy to verify that R4 and R5 are satisfied for the first $m$ nodes
due to \eqref{Eqn_Vt}   and the fact that $\{e_0,\cdots,e_{2^m-1}\}$ is the standard basis.
\hfill$\blacksquare$

\textbf{Proof of Theorem \ref{Thm_4_4}}:
We only prove item (iii) of Theorem \ref{Thm_4_1} hereafter since the other items of Theorems \ref{Thm_4_1} and \ref{Thm_4_2} can be
easily verified.

Given two integers $1\le i\le m$ and $m+1\le j\le 3m$, if $j\equiv i \hspace{-1mm}\pmod m$, then
\begin{equation*}
  \lambda_{i,1}(\lambda_{i,0}-k_{j-m})=\gamma^{i}(\gamma^{i}+2\gamma^i)\ne -\gamma^{2i}=\lambda_{j,0}\lambda_{j,1}
\end{equation*}
since $4\gamma^i\ne 0$; Otherwise, define
$j'=j-lm$ where $lm+1\le j\le (l+1)m$ for $1\le l\le $2, i.e., $1\le j'\ne i\le m$, then we have
\begin{equation*}
  \lambda_{i,0}\lambda_{i,1}=\gamma^{2i}\ne\gamma^{2j'}=\lambda_{j,s}^2\ \mbox{for}\ s=0,1.
\end{equation*}
Thus, item (iii) of Theorem \ref{Thm_4_1} is satisfied.
\hfill$\blacksquare$

\textbf{Proof of Theorem \ref{Thm_1}}:
$\mathcal{C}_3$ has  the MDS property if and only if R1 holds. In what follows,
we only prove it for the case that  $1\le i\le m, m+1\le j\le 2m$. The other cases can be proven similarly as those of Cases 1-2 in the proof of Theorem \ref{Thm_4_1}.

When $1\le i\le m$ and $m+1\le j\le 2m$, if $j=i+m$, by (\ref{B3}) we have
\begin{equation*}
    \left(
       \begin{array}{c}
         V_{i,0} \\
         V_{i,1} \\
       \end{array}
     \right)(A_i-A_j)=\left(
                          \begin{array}{c}
                            \lambda_{i,1}V_{i,1}-\lambda_{j,0}V_{i,0} \\
                            \lambda_{i,0}V_{i,0}- \lambda_{j,1}V_{i,1}\\
                          \end{array}
                        \right),
\end{equation*}
which together with  Lemma \ref{rank} gives
\begin{equation*}
    \mbox{rank}(A_i-A_j)=\alpha\Leftrightarrow\lambda_{i,1}\lambda_{i,0}\neq \lambda_{j,0}\lambda_{j,1};
\end{equation*}
Otherwise,
\begin{equation*}
    \left(
       \begin{array}{c}
         V_{i,j,0,0} \\
V_{i,j,0,1} \\
V_{i,j,1,0} \\
V_{i,j,1,1} \\
       \end{array}
     \right)(A_i-A_j)=\left(
                   \begin{array}{l}
                     \lambda_{i,1}V_{i,j,1,0}-\lambda_{j,0}V_{i,j,0,0} \\
                     \lambda_{i,1}V_{i,j,1,1}-\lambda_{j,1}V_{i,j,0,1} \\
                     \lambda_{i,0}V_{i,j,0,0}-\lambda_{j,0}V_{i,j,1,0} \\
                     \lambda_{i,0}V_{i,j,0,1} -\lambda_{j,1}V_{i,j,1,1}\\
                   \end{array}
                 \right)
\end{equation*}
associated with  Lemma \ref{rank}, which implies that
\begin{equation*}
    \mbox{rank}(A_i-A_j)=\alpha\Leftrightarrow\lambda_{i,0}\lambda_{i,1}\neq\lambda_{j,0}^2,\lambda_{j,1}^2.
\end{equation*}
\hfill$\blacksquare$

\textbf{Proof of Theorem \ref{Thm_2}}:
$\mathcal{C}_3$ is a code with the optimal repair property if and only if R2 and R3 hold.
For $m+1\le i\le 2m$, we have
\begin{eqnarray*}\mbox{rank}\left(\left(
                                                        \begin{array}{c}
                                                          S_i \\
                                                          S_iA_i \\
                                                        \end{array}
                                                      \right)\right)\hspace{-2mm}&=&\hspace{-2mm}\mbox{rank}\left(\left(
                                                                \begin{array}{c}
                                                                  V_{i,0}+t_{i-m}V_{i,1} \\
                                                                  \lambda_{i,0}V_{i,0}+\lambda_{i,1}t_{i-m}V_{i,1} \\
                                                                \end{array}
                                                              \right)\right)\\
&=&\hspace{-2mm}\alpha \\&\Leftrightarrow &\hspace{-2mm}\lambda_{i,0}\neq \lambda_{i,1}.
\end{eqnarray*}
The analysis for the remainder cases are omitted herein since they are similar to those of $\mathcal{C}_1$.
\hfill$\blacksquare$

\textbf{Proof of Theorem \ref{Thm_code1}}:
Since the other items of Theorems \ref{Thm_1} and \ref{Thm_2} can be
easily satisfied, we only verify item (iii) of Theorem \ref{Thm_1} herein.

Given two integers $1\le i\le m$ and  $m+1\le j\le 2m$, define $j'=j-m$. Obviously, $1\le j'\le m$. If $j'=i$,  we have
$ \lambda_{i,0}\lambda_{i,1}=\gamma^{2i}\ne \gamma^{\lfloor\frac{q}{2}\rfloor+2i}=\lambda_{j,0}\lambda_{j,1}$;
Otherwise,
\begin{equation*}
  \frac{\lambda_{j,0}^2}{\lambda_{i,0}\lambda_{i,1}}=\frac{\gamma^{2j'}}{\gamma^{2i}}=\gamma^{2j'-2i}\ne 1,
\end{equation*}
and
\begin{eqnarray*}
  \frac{\lambda_{j,1}^2}{\lambda_{i,0}\lambda_{i,1}}=\gamma^{2\lfloor\frac{q}{2}\rfloor+2j'-2i}
  &=&
  \left\{\begin{array}{cc}
  \gamma^{2j'-2i}, & q ~ \textrm{odd} \\
  \gamma^{2j'-2i+1}, & q ~ \textrm{even}
   \end{array}
  \right.
  \ne 1
\end{eqnarray*}
where we use the facts that $1\le|2j'-2i+1|\le 2m-1\le q-2$ and $\gamma^{l}=1$ if and only if $l\equiv0\,(\bmod\,q-1)$.
Thus, item (iii) of Theorem \ref{Thm_1} is satisfied.
\hfill$\blacksquare$

\textbf{\textbf{Proof of Theorem \ref{Thm_2_1}}}: $\mathcal{C}_4$ has the MDS property if and only if R1 holds.

When $1\leq i<j\leq k$, if $j\ne i+m$, similarly as Case 1 in the proof of Theorem \ref{Thm_4_1}, we have
\begin{equation*}
    \mbox{rank}(A_i-A_j)=\alpha\Leftrightarrow\lambda_{i,0}\lambda_{i,1}\neq \lambda_{j,0}\lambda_{j,1};
\end{equation*}
Otherwise,
by (\ref{B3}) we have
\begin{eqnarray*}
    \left(
       \begin{array}{c}
         V_{i,0} \\
         V_{i,1} \\
       \end{array}
     \right)(A_i-A_j)&=&\left(
       \begin{array}{c}
         \lambda_{i,1}V_{i,1} \\
\lambda_{i,0}V_{i,0} \\
       \end{array}
     \right)-\left(
                  \begin{array}{c}
                    \lambda_{j,1}V_{i,1} \\
                    \lambda_{j,0}V_{i,0} \\
                  \end{array}
                \right)\\&=&\left(
                          \begin{array}{c}
                            (\lambda_{i,1}- \lambda_{j,1})V_{i,1} \\
                            (\lambda_{i,0}-\lambda_{j,0})V_{i,0}\\
                          \end{array}
                        \right),
\end{eqnarray*}
which together with Lemma \ref{rank} implies
\begin{equation*}
    \mbox{rank}(A_i-A_j)=\alpha\Leftrightarrow\lambda_{i,s}\neq \lambda_{j,s}\ \ \mbox{for\ \ }s=0,1.
\end{equation*}
\hfill$\blacksquare$

\textbf{Proof of Theorem \ref{Thm_2_2}}: $\mathcal{C}_4$ is a code with the optimal repair  property if and only if R2 and R3 hold.

For $1\leq i\leq k$, we have
\begin{eqnarray*}
\mbox{rank}\left(\left(
                                                        \begin{array}{c}
                                                          S_i \\
                                                          S_iA_i \\
                                                        \end{array}
                                                      \right)\right)&=&\mbox{rank}\left(\left(
                                                                \begin{array}{l}
                                                                  V_{i,0}+t_{i}V_{i,1} \\
                                                                  \lambda_{i,1}V_{i,1}+t_i\lambda_{i,0}V_{i,0} \\
                                                                \end{array}
                                                              \right)\right)\\
&=&\alpha \\&\Leftrightarrow& \lambda_{i,1}\neq t_i^2\lambda_{i,0}.
\end{eqnarray*}

Similar to Case 2(a) in the proof of Theorem \ref{Thm_4_2}, we can get
\begin{eqnarray*}
&&\mbox{rank}\left(\left(
                 \begin{array}{c}
                   S_i \\
                   S_iA_j \\
                 \end{array}
               \right)\right)=\frac{\alpha}{2}\ \mbox{for\ any\ }1\leq i\neq j\leq k\\ &\Leftrightarrow& \lambda_{i,1}=t_{i+m}^2\lambda_{i,0}, \lambda_{i+m,1}=t_i^2\lambda_{i+m,0}\ \mbox{for\ all\ }1\leq i\leq m.
\end{eqnarray*}
\hfill$\blacksquare$

\section*{Acknowledgement}

The authors would like to thank the Associate Editor
Professor Michael Langberg and the two anonymous referees
for their helpful comments, which have greatly improved the presentation and
quality of this paper.

\begin{IEEEbiographynophoto}{Jie Li}
received the B.S. and M.S. degrees in mathematics from Hubei
University, Wuhan, China, in 2009 and 2012, respectively. He is currently pursuing
Ph.D. degree at Southwest Jiaotong University, Chengdu, China. His research
interest includes coding for distributed storage and sequence design.
\end{IEEEbiographynophoto}

\begin{IEEEbiographynophoto}{Xiaohu Tang} (M'04) received the B.S. degree in applied mathematics from
the Northwest Polytechnic University, Xi'an, China, the M.S. degree in applied
mathematics from the Sichuan University, Chengdu, China, and the Ph.D.
degree in electronic engineering from the Southwest Jiaotong University,
Chengdu, China, in 1992, 1995, and 2001 respectively.

From 2003 to 2004, he was a research associate in the Department of Electrical
and Electronic Engineering, Hong Kong University of Science and Technology.
From 2007 to 2008, he was a visiting professor at University of Ulm,
Germany. Since 2001, he has been in the School of Information Science and Technology,
Southwest Jiaotong University, where he is currently a professor. His research
interests include coding theory, network security, distributed storage and information processing for big data.

Dr. Tang was the recipient of the National excellent Doctoral Dissertation
award in 2003 (China), the Humboldt Research Fellowship in 2007 (Germany),
and the Outstanding Young Scientist Award by NSFC in 2013 (China). He serves as the
Associate Editor of the IEICE Trans on Fundamentals, and Guest Editor/Associate-Editor for special section on sequence design and its application in communications.
\end{IEEEbiographynophoto}

\begin{IEEEbiographynophoto}{Udaya Parampalli} (M'90-SM'12) (aka Parampalli Udaya) obtained his
doctoral degree in Electrical Engineering from Indian institute of Technology
(I.I.T), Kanpur, in 1993. From 1992 to 1996, he worked in Industry as a
Member Research Staff at Central Research Laboratory, Bharat Electronics,
Bangalore. From 1997 to 2000, he was an ARC research associate at the
Department of Mathematics, RMIT University, Melbourne, Australia. Since
February 2000, he has been working at the Department of Computer Science
and Software Engineering, the University of Melbourne, which in 2012 was
merged with the newly formed Department of Computing and Information
Systems. Currently he is an Associate Professor and Reader at the department.
His research interests are in the area of coding theory, cryptography and
sequences over finite fields and rings for communications and information
security.
\end{IEEEbiographynophoto}
\end{document}